\documentclass[twocolumn,showpacs,aps,epsfig]{revtex4}

\usepackage{graphicx}
\usepackage{epstopdf}
\usepackage{latexsym}

\usepackage[center]{subfigure}

\begin{document}

 \newcommand{\bq}{\begin{equation}}
 \newcommand{\eq}{\end{equation}}
 \newcommand{\bqn}{\begin{eqnarray}}
 \newcommand{\eqn}{\end{eqnarray}}
 \newcommand{\nb}{\nonumber}
 \newcommand{\lb}{\label}
\newcommand{\PRL}{Phys. Rev. Lett.}
\newcommand{\PL}{Phys. Lett.}
\newcommand{\PR}{Phys. Rev.}
\newcommand{\CQG}{Class. Quantum Grav.}

\title{Late transient acceleration of the universe in string theory on $S^{1}/Z_{2}$}
\author{Qiang Wu $^{1,2}$}
\email{qiang_wu@baylor.edu}
\author{N.O. Santos $^{3,4,5}$}
\email{nos@cbpf.br and N.O.Santos@qmul.ac.uk}
\author{Pamela Vo $^2$}
\email{pamela_vo@baylor.edu}
\author{Anzhong Wang $^{2,3,6}$}
\email{anzhong_wang@baylor.edu}
\affiliation{$^{1}$  Department of Applied Physics,
Zhejiang University of Technology, Hangzhou 310032,  China 
\\
$^{2}$ GCAP-CASPER, Department of Physics, Baylor University,
Waco, Texas 76798-7316\\
 $^{3}$ LERMA/CNRS-FRE 2460, Universit\'e Pierre et Marie Curie, ERGA, 
Bo\^{\i}te 142, 4 Place Jussieu, 75005 Paris Cedex 05,
France\\
$^{4}$ School of Mathematical Sciences, Queen Mary,
University of London, London E1 4NS, UK\\
$^{5}$ Laborat\'orio Nacional de Computa\c{c}\~{a}o Cient\'{\i}fica, 
25651-070 Petr\'opolis RJ, Brazil\\
$^{6}$ Department of Theoretical Physics, the State University of
Rio de Janeiro, RJ, Brazil}

\date{\today}

\begin{abstract}

Recently, in Gong {\em et al} \cite{GWW07} and Wang and Santos \cite{WS07} it was shown  that 
the effective cosmological 
constant on each of the two orbifold branes can be easily lowered to its current observational 
value, by using the large extra dimensions in the framework of both M-Theory and string theory 
on $S^{1}/Z_{2}$. In this paper, we study the current acceleration of the universe, using the 
formulas developed in \cite{WS07}. We first construct explicitly time-dependent solution 
to the 10-dimensional bulk of the Neveu-Schwarz/Neveu-Schwarz sector, compactified on a 
5-dimensional torus. Then, we write down the generalized Friedmann equations on each of the 
two dynamical branes, and fit the models to the  182 gold supernova Ia data and the BAO 
parameter from SDSS, using both of our MINUIT and Monte-Carlo Markov Chain (MCMC) codes. 
With the best fitting values of the parameters involved as initial conditions, we integrate 
the generalized Friedmann equations 
numerically and find the future evolution of the universe. We find that it depends on the 
choice of the radion potentials $V_{4}^{(I)} \; (I = 1, 2)$ of the branes. In particular, 
when choosing them to be the Goldberger-Wise potentials, $V_{4}^{(I)} = \lambda_{4}^{(I)}
\left(\psi^{2} - {v_{I}}^{2}\right)^{2}$, we find that the current acceleration of the 
universe driven by the effective cosmological constant is only temporary.  Due to the 
effects of the potentials, the universe will be finally in its decelerating expansion phase 
again. We also study the proper distance between the two branes, and find that it remains 
almost constant during the whole future evolution of the universe in all the models 
considered.

\end{abstract}
\pacs{98.80.-k,04.20.Cv,04.70.Dy}
\preprint{arXiv:0804.0620}

\vspace{.7cm}

\pacs{ 03.50.+h, 11.10.Kk, 98.80.Cq, 97.60.-s}

\maketitle

\vspace{1.cm}

\section{Introduction}

One of the long-standing problems in particle physics and cosmology is the cosmological
constant problem: its theoretical expectation values  from quantum field theory
exceed observational limits by $120$ orders of magnitude  \cite{wen}. Even if
such high energies are suppressed by supersymmetry, the electroweak corrections  
are still $56$ orders higher.  This problem was further sharpened by recent  
observations of supernova (SN) Ia, which  reveal the striking discovery that 
our universe has lately been in its accelerated expansion phase  
\cite{agr98}. Cross checks from the cosmic microwave background radiation  
and large scale structure all confirm this unexpected result \cite{Obs}. 

In Einstein's theory of gravity, such an expansion can be achieved by 
a tiny positive cosmological constant. In fact, such a  constant is well 
consistent with all observations carried out so far \cite{SCH07}. 
Therefore,  solving the cosmological constant problem now becomes more urgent than 
ever before. As a matter of fact, it is exactly because of this that a large number 
of ambitious projects 
have been proposed lately to distinguish the cosmological constant from  dynamical
dark energy models \cite{DETF}.

Since the problem is intimately related to quantum gravity, its solution is expected 
to come from quantum gravity, too. At the present, string/M-Theory is our best bet 
for a consistent  quantum theory of gravity, so it is  reasonable to ask what 
string/M-Theory has to say about the cosmological constant. In the string landscape 
\cite{Susk}, it is expected that there are many different vacua with different local 
cosmological constants \cite{BP00}. Using the anthropic principle, 
one may select the low energy vacuum in which we can exist. However, many theorists 
still hope to explain the problem without invoking the existence of ourselves in the 
universe. 

Townsend and Wohlfarth \cite{townsend} considered a time-dependent compactification  
of pure gravity in higher dimensions with hyperbolic internal space to circumvent 
Gibbons' non-go theorem \cite{gibbons}. Their exact solution   exhibits 
a short period of acceleration. The solution is the zero-flux limit 
of spacelike branes \cite{ohta}. If non-zero flux or forms are turned on,  a transient 
acceleration exists for both compact internal hyperbolic and flat 
spaces \cite{wohlfarth}. Other accelerating solutions  by compactifying 
more complicated time-dependent internal spaces can be found  in \cite{string}.  

In the same spirit,  the cosmological constant problem was also
studied in the framework of brane world in 5-dimensional spacetimes \cite{5CC} and 6-dimensional 
supergravity \cite{6CC}. However, it turned out that in the 5--dimensional case hidden 
fine-tunings are required \cite{For00}, while in the 6-dimensional case it is still not 
clear whether loop corrections can be as small as  required \cite{Burg07}.

Recently, we \cite{GWW07} studied the cosmological constant problem   and late 
acceleration  of the universe in the framework of  the Horava-Witten
heterotic M-Theory    on $S^{1}/Z_{2}$ \cite{HW96}. In particular, using the 
Arkani-Hamed-Dimopoulos-Dvali (ADD) mechanism of large extra dimensions \cite{ADD98}, 
we showed explicitly that the effective cosmological constant 
on each of the two orbifold branes can be easily lowered to its current observational 
value. Applying it to cosmology,   we further found that the domination of the
effective cosmological constant is only temporary.  Due to the interaction of the bulk 
and the brane, the universe will be in its decelerating expansion phase again in the 
future, whereby all problems connected with string cosmology  \cite{KS00}
are resolved. Such studies were also generalized to string theory, and found 
that the ADD mechanism can be used in the same way 
to solve the cosmological constant problem  \cite{WS07}.  Therefore, the ADD mechanism 
for solving both the cosmological constant problem and the  hierarchy  problem 
is a built-in mechanism in the brane world of string/M-Theory.

In this paper, we apply the theory developed in \cite{WS07} in the framework of
string theory on $S^{1}/Z_{2}$ to cosmology, and study the current acceleration of the 
universe. In particular, in Sec. II we  first give a brief review of the theory, 
and then write down the field equations both in the bulk and on the two orbifold branes.
In Sec. III, we present a particular time-dependent solution to these equations,
and  study its local and global properties. Then, we  write down explicitly the
generalized Friedmann equations on each of the two branes for any radion potentials 
$V_{4}^{(I)} \; (I = 1, 2)$ of the branes. Depending on the choice of $V_{4}^{(I)}$'s,
the future evolution of the universe is different. We study two different cases.
We  fit these models to the  182 
gold supernova Ia data \cite{Riess06} and BAO parameter from SDSS \cite{SDSS}, using 
both of our MINUIT  \cite{Gong}
and Monte-Carlo Markov Chain (MCMC)  \cite{GWW07b} codes. With these best fitting values 
as the initial condition, we integrate numerically the field equations on the branes to 
find the future evolution of the universe. In the latter case,  we show that the current 
acceleration of the universe driven by the effective cosmological constant is only 
temporary.  Due to   the effects of the potentials, the universe will be in its decelerating 
expansion phase again. We also study the proper distance between the 
two orbifold branes, and find that it remains almost constant during the whole 
future evolution of the universe in all these models. In Sec. IV, we summarize our main 
results and present some concluding remarks.

Before turning to the next section, we would like to note that Sahni and Shtanov \cite{SS03}
found that transient acceleration of the universe happened in the DGP brane model \cite{DGP},
too.

\section{Brany cosmology of string theory on $S^{1}/Z_{2}$}

\renewcommand{\theequation}{2.\arabic{equation}}
\setcounter{equation}{0}

To begin with, in this section we give a brief review on the cosmological 
models of orbifold branes developed  in  \cite{WS07}.  In this paper we shall restrict
ourselves directly to the case $D = d = 5$. For the case with arbitrary $D$ and $d$,
we refer readers to \cite{WS07}.

\subsection{The Model}

For the toroidal compactification 
of the Neveu-Schwarz/Neveu-Schwarz (NS-NS) sector  
in (5+5) dimensions, $\hat{M}_{10} = M_{5}\times {\cal{T}}_{5}$, where  ${\cal{T}}_{5}$
is a 5-dimensional torus, the action takes the form \cite{LWC00,BW06},
\bqn
\lb{2.1}
\hat{S}_{10} &=& - \frac{1}{2\kappa^{2}_{10}}
\int{d^{10}x\sqrt{\left|\hat{g}_{10}\right|}  e^{-\hat{\Phi}} \left\{
{\hat{R}}_{10}[\hat{g}]\right.}\nb\\
& & \left. + \hat{g}^{AB}\left(\hat{\nabla}_{A}\hat{\Phi}\right)
\left(\hat{\nabla}_{B}\hat{\Phi}\right) - \frac{1}{12}{\hat{H}}^{2}\right\},
\eqn
where $\hat{\nabla}_{A}$ denotes the covariant derivative with respect to $\hat{g}^{AB}$
with $A, B = 0, 1, ..., 9$, and $\hat{\Phi}$ is the dilaton field. The NS three-form 
field $\hat{H}_{ABC}$ is defined as
\bq
\lb{2.1a}
\hat{H}_{ABC} = 3 \partial_{[A}\hat{B}_{BC]},
\eq
where  the square brackets imply total antisymmetrization over all indices. 
The $10$-dimensional spacetimes to be considered are described by the metric,
\bqn
\lb{2.3}
d{\hat{s}}^{2}_{10} &=& \hat{g}_{AB} dx^{A}dx^{B} \nb\\
&=&
   \tilde{g}_{ab}\left(x^{c}\right) dx^{a}dx^{b}  +   h_{ij}\left(x^{c}\right)dz^{i} dz^{j},
\eqn
where  $\tilde{g}_{ab}$ is the metric on $M_{5}$, parametrized by the coordinates $x^{a}$
with $a,b, c = 0, 1, ..., 4$, and $h_{ij}$ is the metric on the compact space ${\cal{T}}_{5}$
with periodic coordinates $z^{i}$, where $i,j = 5, 6, ..., 9$. 

By assuming that all the matter fields  are functions of
$x^{a}$ only, it can be shown that  the effective $5-$dimensional action is given by,
\bqn
\lb{2.7}
S_{5} &=& - \frac{1}{2\kappa^{2}_{5}}
\int{d^{5}x\sqrt{\left|\tilde{g}_{5}\right|} e^{-\tilde{\phi}}
\left\{\tilde{R}_{5}[\tilde{g}] + \left(\tilde{\nabla}_{a}\tilde{\phi}\right)
\left(\tilde{\nabla}^{a}\tilde{\phi}\right)\right.}\nb\\
& & + \frac{1}{4} \left(\tilde{\nabla}_{a}h^{ij}\right)\left(\tilde{\nabla}^{a}h_{ij}\right)
- \frac{1}{12}\tilde{H}_{abc}\tilde{H}^{abc}\nb\\
& & \left. - \frac{1}{4} h^{ik}h^{jl}\left(\tilde{\nabla}_{a}B_{ij}\right)
\left(\tilde{\nabla}^{a}B_{kl}\right)\right\},
\eqn
where
\bqn
\lb{2.8a}
\tilde{\phi} &=& \hat{\Phi} - \frac{1}{2}\ln\left|h\right|,\\
\lb{2.8b}
\kappa^{2}_{5} &\equiv& \frac{\kappa^{2}_{10}}{V_{0}},
\eqn
with the $5-$dimensional internal volume given by
\bq
\lb{2.9}
V\left(x^{a}\right) \equiv \int{d^{5}z \sqrt{|h|}} =  |h|^{1/2} V_{0}.
\eq
Note that in writing the action (\ref{2.7}) we had assumed that the flux is block diagonal,
\bq
\lb{2.9a}
\left(\hat{B}_{CD}\right) = \left(\matrix{\tilde{B}_{ab} & 0\cr
0 & B_{ij}\cr}\right).
\eq
The action (\ref{2.7}) is usually referred to as written in the string frame.
To go to the Einstein frame, we make the following conformal transformations,
\bqn
\lb{2.10}
g_{ab} &=& \Omega^{2}\tilde{g}_{ab},\nb\\
\Omega^{2} &=& \exp\left(-\frac{2}{3}\tilde{\phi}\right),\nb\\
\phi &=& \sqrt{\frac{2}{3}} \; \tilde{\phi}.
\eqn
Then, the action (\ref{2.7}) takes the form
\bqn
\lb{2.11}
S_{5}^{(E)} &=& -\frac{1}{2\kappa^{2}_{5}}
\int{d^{5}x\sqrt{\left|{g}_{5}\right|}  
\left\{{R}_{5}[{g}] - \frac{1}{2}\left(\nabla\phi\right)^{2}\right.}\nb\\
& &  + \frac{1}{4} \left({\nabla}_{a}h^{ij}\right)\left({\nabla}^{a}h_{ij}\right)\nb\\
& & - \frac{1}{12} e^{- \sqrt{\frac{8}{3}}\phi}H_{abc}H^{abc}\nb\\
& & \left. - \frac{1}{4} h^{ik}h^{jl}\left({\nabla}_{a}B_{ij}\right)
\left({\nabla}^{a}B_{kl}\right)\right\},
\eqn
where $\nabla_{a}$ denotes the covariant derivative with respect to $g_{ab}$. It should be
noted that, since the definition of the three-form $\hat{H}_{ABC}$ given by (\ref{2.1a}) is 
independent of the metric, it is conformally invariant. In particular, we have
$H_{abc} = \tilde{H}_{abc}$ and $B_{ab} = \tilde{B}_{ab}$. However, we do have
\bqn
\lb{2.11a}
& & {H}^{abc} = g^{ad}g^{be}g^{cf}H_{def} = \Omega^{-6}\tilde{H}^{abc},\nb\\
& & H_{abc}H^{abc} = \Omega^{-6}\tilde{H}_{abc}\tilde{H}^{abc}. 
\eqn

Considering the addition of a potential term  \cite{BW06},  in the string frame we have
\bq
\lb{2.12}
\hat{S}^{m}_{10} = - \int{d^{10}x \sqrt{\left|\hat{g}_{10}\right|} V_{10}^{s}}.
\eq
Then, after the dimensional reduction we find
\bq
\lb{2.13}
{S}_{5, m} = - V_{0} \int{d^{5}x \sqrt{\left|\tilde{g}_{5}\right|} \; {|h|}^{1/2} V_{10}^{s}},
\eq
where
\bq
\lb{2.14}
\tilde{g}_{5} = \exp\left(\sqrt{\frac{50}{3}}\; \phi\right)\; g_{5}.
\eq
Changed to the Einstein frame, the action (\ref{2.13}) finally takes the form,
\bq
\lb{2.13a}
{S}_{5, m}^{(E)} = - \frac{1}{2\kappa^{2}_{5}} \int{d^{5}x \sqrt{\left|{g}_{5}\right|}   V_{5}},
\eq
where
\bq
\lb{2.14a}
V_{5} \equiv 2\kappa^{2}_{5}V_{0}V_{10}^{s} 
\exp\left(\frac{5}{\sqrt{6}}\;\phi\right)\; |h|^{1/2}.
\eq
If we further assume that
\bqn
\lb{2.15}
h_{ij} &=& - \exp\left(\sqrt{\frac{2}{5}}\; \psi\right) \delta_{ij},\nb\\ 
h^{ij} &=& - \exp\left(-\sqrt{\frac{2}{5}}\; \psi\right) \delta^{ij},
\eqn
we find that
\bqn
\lb{2.16}
S_{5}^{(E)} + S_{5, m}^{(E)} &=& - \frac{1}{2\kappa^{2}_{5}}
\int{d^{5}x\sqrt{\left|{g}_{5}\right|}  
\left\{{R}_{5}[{g}] \right.}\nb\\
& & - \frac{1}{2}\left(\left(\nabla\phi\right)^{2} + \left(\nabla\psi\right)^{2}
- 2V_{5}\right)\nb\\
& &  - \frac{1}{4} e^{- \sqrt{\frac{8}{5}}\; \psi}
 \delta^{ik}\delta^{jl}\left({\nabla}_{a}B_{ij}\right)
\left({\nabla}^{a}B_{kl}\right)\nb\\
& & \left. - \frac{1}{12} e^{- \sqrt{\frac{8}{3}}\; \phi}H_{abc}H^{abc}\right\},
\eqn
where the effective $5-$dimensional potential (\ref{2.14}) now becomes
\bq
\lb{2.17}
V_{5} \equiv V_{(5)}^{0} \exp\left(\frac{5}{\sqrt{6}}\;\phi 
 + \sqrt{\frac{5}{2}}\;\psi\right),
\eq
where  $V_{(5)}^{0} \equiv 2\kappa^{2}_{5} V_{0} V^{s}_{10}$.

To study orbifold branes, we consider the brane actions,
\bqn
\lb{3.1}
S^{(I)}_{4, m} &=& -  \int_{M^{(I)}_{4}}{\sqrt{\left|g^{(I)}_{4}\right|}
\left(\epsilon_{I}V^{(I)}_{4}(\phi, \psi) + g^{(I)}_{s}\right)d^{4}\xi_{(I)}} \nb\\
& & + \int_{M^{(I)}_{4}}{d^{4}\xi_{(I)}\sqrt{\left|g^{(I)}_{4}\right|}}\nb\\
& & \times {\cal{L}}^{(I)}_{4,m}\left(\phi,\psi, B, \chi\right),
\eqn
where $I = 1, 2$, $\; V^{(I)}_{4}(\phi, \psi)$  denotes the potential of the scalar fields 
$\phi$ and $\psi$, 
and $\xi_{(I)}^{\mu}$'s are the intrinsic coordinates of the I-th brane with $\mu, \nu = 0, 1, 
2, 3$, and $\epsilon_{1} = - \epsilon_{2} = 1$. $\chi$  denotes collectively the matter fields,
and $g^{(I)}_{s}$ is a constant, which is related to the four-dimensional Newtonian constant
via the relation given by Eq.(\ref{3.17}) below.
The variation of the total action,
\bq
\lb{action}
S_{total} = S_{5}^{(E)} + S_{5, m}^{(E)} + \sum_{I = 1}^{2}{S^{(I)}_{4, m}},
\eq
with respect to the metric $g_{ab}$ yields the field equations,
\bqn
\lb{3.3}
G^{(5)}_{ab} &=& \kappa^{2}_{5}T^{(5)}_{ab} + \kappa^{2}_{5} 
\sum^{2}_{I=1}{{\cal{T}}^{(I)}_{\mu\nu} e^{(I, \; \mu)}_{a}e^{(I, \; \nu)}_{b}}\nb\\
& &  \times \sqrt{\left|\frac{g^{(I)}_{4}}{g_{5}}\right|} \delta\left(\Phi_{I}\right),
\eqn
where  $\delta(x)$ denotes the Dirac delta function normalized in the sense of \cite{LMW01},
and the two branes are localized on the surfaces,
\bq
\lb{3.3d}
\Phi_{I}\left(x^{a}\right)  = 0.
\eq 
The energy-momentum tensors $T^{(5)}_{ab}$ and ${\cal{T}}^{(I)}_{\mu\nu}$ are given by
\bqn
\lb{3.3a}
\kappa^{2}_{5}T^{(5)}_{ab} &\equiv & \frac{1}{2}
\left[\left(\nabla_{a}\phi\right)\left(\nabla_{b}\phi\right)
 + \left(\nabla_{a}\psi\right)\left(\nabla_{b}\psi\right)\right.\nb\\
& & + \frac{1}{2}e^{-\sqrt{\frac{8}{5}}\; \psi}  
\left(\nabla_{a}B^{ij}\right)\left(\nabla_{b}B_{ij}\right)\nb\\
& & \left. + \frac{1}{2}e^{\sqrt{\frac{8}{3}}\; \phi} H_{acd}H_{b}^{\;\; cd}\right]\nb\\
& & - \frac{1}{4}g_{ab}
\left[\left(\nabla\phi\right)^{2} + 
\left(\nabla\psi\right)^{2}  - 2V_{5}\right. \nb\\
& &   \frac{1}{2}e^{-\sqrt{\frac{8}{5}}\; \psi}  
\left(\nabla_{c}B^{ij}\right)\left(\nabla^{c}B_{ij}\right)\nb\\
& & \left. + \frac{1}{6}e^{\sqrt{\frac{8}{3}}\; \phi} H_{cde}H^{cde}\right],\nb\\
{\cal{T}}^{(I)}_{\mu\nu} &\equiv& {\tau}^{(I)}_{\mu\nu} 
+ \left( g^{(I)}_{s} + \tau^{(I)}_{(\phi,\psi)}\right) g_{\mu\nu}^{(I)},\nb\\
{\tau}^{(I)}_{\mu\nu} &\equiv& 2\frac{\delta{\cal{L}}^{(I)}_{4,m}}
{\delta{g^{(I)\; \mu\nu}}} -  g^{(I)}_{\mu\nu}{\cal{L}}^{(I)}_{4,m},
\eqn
where $B^{ij} \equiv \delta^{ik}\delta^{jl}B_{kl}$, 
\bqn
\lb{3.3b}
\tau^{(I)}_{(\phi,\psi)} &\equiv& \epsilon_{I}V^{(I)}_{4}(\phi,\psi), \nb\\
e^{(I)\; a}_{(\mu)} &\equiv& \frac{\partial x^{a}}{\partial \xi^{\mu}_{(I)}},\nb\\
e^{(I,\; \mu)}_{a} &\equiv& g_{ab} g^{(I)\; \mu\nu} e^{(I)\; b}_{(\nu)},  
\eqn
and $g_{\mu\nu}^{(I)}$ is the reduced metric on the I-th brane, defined as
\bq
\lb{3.3c}
g_{\mu\nu}^{(I)} \equiv \left. g_{ab} e^{(I)a}_{(\mu)} e^{(I)b}_{(\nu)}\right|_{M^{(I)}_{4}}. 
\eq

Variation of the total action Eq.(\ref{action}) 
with respect to $\phi, \; \psi$ and $B$, respectively, yields the following
equations of the matter fields,
\bqn
\lb{3.ea}
\Box\phi &=& - \frac{\partial{V_{5}}}{\partial{\phi}} 
- \frac{1}{12}\sqrt{\frac{8}{3}}e^{-\sqrt{\frac{8}{3}}\; \phi}H_{abc}H^{abc}\nb\\
& & - 2\kappa_{5}^{2}\sum^{2}_{I=1}{\left(\epsilon_{I}\frac{\partial{V_{4}^{(I)}}}
{\partial{\phi}} + \sigma^{(I)}_{\phi}\right)}\nb\\
& & \times\sqrt{\left|\frac{g^{(I)}_{4}}{g_{5}}\right|} \delta\left(\Phi_{I}\right),\\
\lb{3.eb}
\Box\psi &=& - \frac{\partial{V_{5}}}{\partial{\psi}} 
- \sqrt{\frac{1}{10}} e^{-\sqrt{\frac{8}{5}}\; \psi}\left(\nabla_{a}B^{ij}\right)
\left(\nabla^{a}B_{ij}\right)\nb\\
& & - 2\kappa_{5}^{2}\sum^{2}_{I=1}{\left(\epsilon_{I}\frac{\partial{V_{4}^{(I)}}}
{\partial{\psi}} + \sigma^{(I)}_{\psi}\right)}\nb\\
& & \times\sqrt{\left|\frac{g^{(I)}_{4}}{g_{5}}\right|} \delta\left(\Phi_{I}\right),\\
\lb{3.ec}
\Box B_{ij} &=& \sqrt{\frac{8}{5}}\left(\nabla_{a}\psi\right)\left(\nabla^{a}B_{ij}\right)\nb\\
& & -  \sum^{2}_{I=1}{\Psi^{(I)}_{ij}
\sqrt{\left|\frac{g^{(I)}_{4}}{g_{5}}\right|} \delta\left(\Phi_{I}\right)},\\
\lb{3.ed}
\nabla^{c}H_{cab} &=& \sqrt{\frac{8}{3}}\; H_{cab}\nabla^{c}\phi\nb\\
& & -  \sum^{2}_{I=1}{\Phi^{(I)}_{ab}
\sqrt{\left|\frac{g^{(I)}_{4}}{g_{5}}\right|} \delta\left(\Phi_{I}\right)},
\eqn
where $\Box \equiv g^{ab}\nabla_{a}\nabla_{b}$, and
\bqn
\lb{3.ee}
\sigma_{\phi}^{(I)} &\equiv& -  \frac{\delta{\cal{L}}^{(I)}_{4,m}}{\delta\phi},\nb\\
\sigma_{\psi}^{(I)} &\equiv& -  \frac{\delta{\cal{L}}^{(I)}_{4,m}}{\delta\psi},\nb\\
\Psi^{(I)}_{ij} &\equiv& - 4\kappa^{2}_{5} e^{\sqrt{\frac{8}{5}}\; \psi}
\frac{\delta{\cal{L}}^{(I)}_{4,m}}{\delta{B^{ij}}},\nb\\
\Phi^{(I)}_{ab} &\equiv& - 4\kappa^{2}_{5} e^{\sqrt{\frac{8}{3}}\; \phi}
\frac{\delta{\cal{L}}^{(I)}_{4,m}}{\delta{B^{ab}}}.
\eqn

To write down the field equations on the branes, one can first
express the delta function part of $G^{(5)}_{ab}$ in terms of the discontinuities of
the first derivatives of the metric coefficients, and then equal the delta function parts 
of the two sides of Eq.(\ref{3.3}), as shown systematically in \cite{WCS07}.
The other way is to use the Gauss-Codacci equations to write the $(4)$-dimensional 
Einstein tensor as \cite{SMS},
 \bq
 \lb{3.12}
 G^{(4)}_{\mu\nu} = {\cal{G}}^{(5)}_{\mu\nu} + E^{(5)}_{\mu\nu}
 + {\cal{F}}^{(4)}_{\mu\nu},
 \eq
 where
 \bqn
 \lb{3.13}
 {\cal{G}}_{\mu\nu}^{(5)} &\equiv&  \frac{2}{3}
\left\{G_{ab}^{(5)}e^{a}_{(\mu)} e^{b}_{(\nu)} \right.\nb\\
& & \left.
- \left[G_{ab}n^{a}n^{b} + \frac{1}{4} G^{(5)}\right]g_{\mu\nu}\right\}, \nb\\
E^{(5)}_{\mu\nu} &\equiv& C_{abcd}^{(5)}n^{a}e^{b}_{(\mu)}n^{c}e^{d}_{(\nu)},\nb\\
{\cal{F}}^{(4)}_{\mu\nu} &\equiv&  
 K_{\mu\lambda}K^{\lambda}_{\nu} - KK_{\mu\nu} \nb\\
& & - \frac{1}{2}g_{\mu\nu}\left(K_{\alpha\beta}K^{\alpha\beta} 
    - K^{2}\right),
\eqn
where $n^{a}$ denotes the normal vector to the brane, $G^{(5)}
\equiv g^{ ab} G^{(5)}_{ab}$, and $C_{abcd}^{(5)}$ the Weyl tensor. 
The extrinsic curvature $K_{\mu\nu}$ is defined as
\bq
\lb{3.13a}
K_{\mu\nu} \equiv e^{a}_{(\mu)}e^{b}_{(\nu)}\nabla_{a}n_{b}.
\eq
A crucial step of this approach is the Lanczos equations \cite{Lan22},
\bq
\lb{3.4}
\left[K_{\mu\nu}^{(I)}\right]^{-} - g_{\mu\nu}^{(I)} \left[K^{(I)}\right]^{-} 
= - \kappa^{2}_{5}{\cal{T}}_{\mu\nu} ^{(I)},
\eq
where 
\bqn
\lb{3.5}
\left[K_{\mu\nu}^{(I)}\right]^{-} &\equiv& {\rm lim}_{\Phi_{I} \rightarrow 0^{+}}
K^{(I)\; +}_{\mu\nu} - {\rm lim}_{\Phi_{I} \rightarrow 0^{-}}
K^{(I)\; -}_{\mu\nu},\nb\\
\left[K^{(I)}\right]^{-} &\equiv& g^{(I)\; \mu\nu}\left[K_{\mu\nu}^{(I)}\right]^{-}.
\eqn

Assuming that the branes have $Z_{2}$ symmetry, we can express the intrinsic
curvatures $K^{(I)}_{\mu\nu}$ in terms of the effective energy-momentum tensor
${\cal{T}}_{\mu\nu} ^{(I)}$ through the Lanczos equations (\ref{3.4}). Then, 
we find that
 $ G^{(4)}_{\mu\nu}$ given by Eq.(\ref{3.12}) can be cast in the form,
 \bqn
 \lb{3.15}
 G^{(4)}_{\mu\nu} &=& {\cal{G}}^{(5)}_{\mu\nu} + E^{(5)}_{\mu\nu}
 + {\cal{E}}_{\mu\nu}^{(4)}\nb\\
 & & + \kappa^{2}_{4}\tau_{\mu\nu} + \Lambda g_{\mu\nu}
 + \kappa^{4}_{5}\pi_{\mu\nu},
 \eqn
 where
 \bqn
\lb{3.16}
\pi_{\mu\nu} &\equiv& \frac{1}{4}\left\{\tau_{\mu\lambda}\tau^{\lambda}_{\nu}
-  \frac{1}{3}\tau \tau_{\mu\nu}\right.\nb\\
& & \left. 
 - \frac{1}{2}g_{\mu\nu}\left(\tau^{\alpha\beta} \tau_{\alpha\beta}
 - \frac{1}{3}\tau^{2}\right)\right\},\nb\\
 {\cal{E}}_{\mu\nu}^{(4)} &\equiv& \frac{\kappa^{4}_{5}}{6}\tau_{(\phi,\psi)}
 \left[\tau_{\mu\nu}  
   + \left(g_{s} + \frac{1}{2}\tau_{(\phi,\psi)}\right)g_{\mu\nu}\right],\nb\\
\eqn
and 
\bqn
\lb{3.17}
\kappa^{2}_{4} &=& \frac{1}{6}g_{s}\kappa^{4}_{5},\nb\\
\Lambda  &=&  \frac{1}{12}g_{s}^{2}\kappa^{4}_{5}.
\eqn
 For a perfect fluid,
\bq
\lb{3.18}
\tau_{\mu\nu} = \left(\rho + p\right)u_{\mu}u_{\nu} - p g_{\mu\nu},
\eq
where $u_{\mu}$ is the four-velocity of the fluid, we find that 
\bq
\lb{3.19}
\pi_{\mu\nu} = \frac{\rho}{6} \left[\left(\rho + p\right)u_{\mu}u_{\nu} 
- \left(p + \frac{1}{2}\rho\right)g_{\mu\nu}\right].
\eq
Note that in writing Eqs.(\ref{3.15})-(\ref{3.19}), without causing any confusion, we had dropped
the super indices $(I)$.

In the rest of this paper, we shall turn off the flux, i.e., $\hat{B}_{CD} = 0$, which is consistent 
with the field equations, provided that $\Psi^{(I)}_{ij} = 0$ and $ \Phi^{(I)}_{ab} = 0$. 

\subsection{The General Metric of the Five-Dimensional Spacetimes}

 Since we shall apply such spacetimes to cosmology,
let us first consider the embedding of a 3-dimensional spatial space that is 
homogeneous, isotropic, and independent of time. It is not difficult to 
show that such a space must have a constant curvature  
and its metric   takes the form \cite{Wolf},
\bq 
\lb{4.1}
d\Sigma^{2}_{k} 
= \frac{dr^{2}}{1 - k r^{2}} + r^{2}\left(d\theta^{2} + \sin^{2}\theta d\phi^{2}\right),
\eq
where the constant $k$  represents the curvature of the 3-space,
and can be positive, negative or zero. Without loss of generality, we shall choose 
coordinates such that $k = 0, \pm 1$. Then, one can see that the most general metric 
for the five-dimensional spacetime must take the
form,
\bq
\lb{4.2}
ds^{2}_{5} = g_{ab} dx^{a}dx^{b}  
 =g_{MN} dx^{M}dx^{N} - e^{2\omega\left(x^{N}\right)}d\Sigma^{2}_{k},
\eq
where  $ M, N = 0, 1$.  Clearly, the metric (\ref{4.2}) is invariant under the coordinate 
transformations,
\bq
\lb{4.3}
{x'}^{N} = f^{N}\left({x}^{M}\right).
\eq
Using  these two degrees of freedom, without loss of generality, we can always set
\bq
\lb{4.3a}
g_{00} = g_{11}, \;\;\; g_{01} = 0, 
\eq
so that the five-dimensional metric finally takes the form, 
\bq
\lb{4.4}
ds^{2}_{5} =  e^{2\sigma(t,y)}\left(dt^{2} - dy^{2}\right)  - e^{2\omega(t,y)}d\Sigma^{2}_{k}.
\eq 
It should be noted that metric (\ref{4.4}) is still subjected to the gauge freedom,
\bq
\lb{4.5}
t = f(t' + y') + g(t' - y'), \;\;\; y = f(t' + y') - g(t' - y'),
\eq
where $f(t' + y')$ and $g(t' - y')$ are arbitrary functions of their indicated
arguments. 

It is also interesting to note that in \cite{WCS07} a different gauge was
used. Instead of setting $g_{00} = g_{11}$ it was chosen that the two branes are
comoving with the coordinates, so that they are located on two fixed hypersurfaces
$ y = 0, \; y_{c}$. For details, see \cite{WCS07}.

\subsection{The Field Equations Outside  the Two Orbifold Branes}

The non-vanishing components of the Ricci tensor outside of 
the two branes are given by
\bqn
\lb{4.6}
R^{(5)}_{tt} &=& \sigma_{,yy} +   3\sigma_{,y}\omega_{,y} 
       -\left[\sigma_{,tt} + 3\omega_{,tt} + 3\omega_{,t}\left(\omega_{,t} 
                 - \sigma_{,t}\right)\right],\nb\\
R^{(5)}_{ty} &=& - 3\left[\omega_{,ty}  + \omega_{,t} \omega_{,y}  
                 - \left(\sigma_{,t}\omega_{,y} + \sigma_{,y}\omega_{,t}\right)\right],\nb\\
R^{(5)}_{yy} &=&  \sigma_{,tt} +   3\sigma_{,t}\omega_{,t} 
                  -   \left[\sigma_{,yy} + 3\omega_{,yy} 
	          + 3\omega_{,y}\left(\omega_{,y} - \sigma_{,y}\right)\right], \nb\\  
R^{(5)}_{mn} &=&  -e^{-2\sigma}g_{mn}\left\{\omega_{,tt} + 3{\omega_{,t}}^{2}
                   - \left(\omega_{,yy} + 3{\omega_{,y}}^{2}\right)\right.\nb\\
	     & & \left. + 2k e^{2(\sigma - \omega)}\right\},    
\eqn
where now $m, \; n = r, \theta, \varphi$, $\; \sigma_{,t} \equiv \partial\sigma/\partial t$
and so on. Then, it can be shown that outside of the two branes
the field equations  have four independent components, which can be cast into the form,
\bqn
\lb{4.6a}
& & \omega_{,tt} + \omega_{,t}\left(\omega_{,t} - 2\sigma_{,t}\right) 
    + \omega_{,yy} + \omega_{,y}\left(\omega_{,y} - 2\sigma_{,y}\right)   \nb\\
    & & \;\;\;\;\;\;\;\;\;\;\;\;\;\;\; 
        = - \frac{1}{6}\left[\left({\phi_{,t}}^{2} + {\phi_{,y}}^{2}\right)
	  + \left({\psi_{,t}}^{2} + {\psi_{,y}}^{2}\right)\right],\\
\lb{4.6b}
& & 2\sigma_{,tt} + \omega_{,tt} - 3 {\omega_{,t}}^{2}
- \left(2\sigma_{,yy} + \omega_{,yy} - 3 {\omega_{,y}}^{2}\right) - 4ke^{2(\sigma-\omega)}  \nb\\
    & & \;\;\;\;\;\;\;\;\;\;\;\;\;\;\; 
       = - \frac{1}{2}\left[\left({\phi_{,t}}^{2} - {\phi_{,y}}^{2}\right)
	  + \left({\psi_{,t}}^{2} - {\psi_{,y}}^{2}\right)\right],\\
\lb{4.6c}
& &  \omega_{,ty} + \omega_{,t} \omega_{,y} 
    -  \left(\sigma_{,t}\omega_{,y} + \sigma_{,y}\omega_{,t}\right) \nb\\
    & & \;\;\;\;\;\;\;\;\;\;\;\;\;\;\; 
       = - \frac{1}{6}\left(\phi_{,t} \phi_{,y} + \psi_{,t}\psi_{,y}\right),\\
\lb{4.6d}
& & \omega_{,tt} + 3{\omega_{,t}}^{2} - \left(\omega_{,yy} + 3{\omega_{,y}}^{2}\right)
     + 2ke^{2(\sigma-\omega)} \nb\\
    & & \;\;\;\;\;\;\;\;\;\;\;\;\;\;\; 
        =  \frac{1}{3}e^{2\sigma} V_{5},
\eqn
where $V_{5}$ is given by Eq.(\ref{2.17}).  On the other hand, the Klein-Gordon
equations   (\ref{3.ea}) and  (\ref{3.eb}) outside the two branes take the form,
\bqn
\lb{4.8a}
& & \phi_{,tt} + 3\phi_{,t}\omega_{,t} - \left(\phi_{,yy} + 3\phi_{,y}\omega_{,y}\right)\nb\\
& & \;\;\;\;\;\;\;\;\;\;\;\;\;\;\;   =  - \frac{5}{\sqrt{6}}\;  V_{5}  e^{2\sigma},\\
\lb{4.8b}
& & \psi_{,tt} + 3\psi_{,t}\omega_{,t} - \left(\psi_{,yy} + 3\psi_{,y}\omega_{,y}\right)\nb\\
& & \;\;\;\;\;\;\;\;\;\;\;\;\;\;\;   =  - \sqrt{\frac{5}{2}}\;  V_{5}  e^{2\sigma}.
\eqn

\subsection{The Field Equations on  the Two Orbifold Branes}

Eqs.(\ref{4.6}) - (\ref{4.8b}) are the field equations that are valid in between the two orbifold branes,
$y_{2}(t_{2}) < y < y_{1}(t_{1})$, where $y = y_{I}(t_{I})$ denote the locations of
the two branes. The proper distance between the two branes is given by,
\bq
\lb{4.9a}
{\cal{D}} \equiv \int_{y_{2}}^{y_{1}}{\sqrt{-g_{yy}} dy}.
\eq
On each of the two branes, the metric reduces to
\bq
\lb{4.9}
\left. ds^{2}_{5}\right|_{M^{(I)}_{4}} = g^{(I)}_{\mu\nu}d\xi_{(I)}^{\mu}d\xi_{(I)}^{\nu}
= d\tau_{I}^{2} - a^{2}\left(\tau_{I}\right)d\Sigma^{2}_{k},
\eq
where $\xi^{\mu}_{(I)} \equiv \left\{\tau_{I}, r, \theta, \varphi\right\}$, and
$\tau_{I}$ denotes the proper time of the I-th brane, defined by
\bqn
\lb{4.10}
d\tau_{I} &=& e^{\sigma}\sqrt{1 - \left(\frac{\dot{y}_{I}}{\dot{t}_{I}}\right)^{2}}\; dt_{I},\nb\\
a\left(\tau_{I}\right) &\equiv& \exp\left\{\omega\left[t_{I}(\tau_{I}), y_{I}(\tau_{I})\right]\right\},
\eqn
with $\dot{y}_{I} \equiv d{y}_{I}/d\tau_{I}$, etc. For the sake of simplicity and without causing any
confusion, from now on we shall drop all the indices ``I", unless some specific attention is 
needed. Then, the normal vector $n_{a}$ and the tangential vectors $e^{a}_{(\mu)}$ are
given, respectively, by
\bqn
\lb{4.11}
n_{a} &=& \epsilon e^{2\sigma}\left(- \dot{y}\delta^{t}_{a} +  \dot{t}\delta^{y}_{a}\right),\nb\\
n^{a} &=& - \epsilon \left(\dot{y}\delta^{a}_{t} +  \dot{t}\delta^{a}_{y}\right),\nb\\
e^{a}_{(\tau)} &=&  \dot{t}\delta^{a}_{t} +  \dot{y}\delta^{a}_{y},
\;\;\; e^{a}_{(r)} = \delta^{a}_{r},\nb\\ 
e^{a}_{(\theta)} &=& \delta^{a}_{\theta},\;\;\;
e^{a}_{(\varphi)} = \delta^{a}_{\varphi},
\eqn
where $\epsilon = \pm 1$. When $\epsilon = +1$, the normal vector $n^{a}$ points toward the increasing
direction of $y$, and when $\epsilon = - 1$, it points toward the decreasing
direction of $y$. Then, the four-dimensional field equations on each of the two branes
take the form,
\bqn
\lb{4.14a}
 H^{2} &+& \frac{k}{a^{2}} = \frac{8\pi G}{3}\left(\rho + \tau_{(\phi, \psi)}\right) 
     + \frac{1}{3}\Lambda 
      + \frac{1}{3}{\cal{G}}^{(5)}_{\tau} + E^{(5)}\nb\\
       & &  
       + \frac{2\pi G}{3\rho_{\Lambda}}\left(\rho + \tau_{(\phi, \psi)}\right)^{2},\\
\lb{4.14b}
\frac{\ddot{a}}{a}   &=&  - \frac{4\pi G}{3}\left(\rho +3p - 2 \tau_{(\phi, \psi)}\right) 
      + \frac{1}{3}\Lambda
      \nb\\
      & &  - E^{(5)} - \frac{1}{6}\left({\cal{G}}^{(5)}_{\tau} + 3{\cal{G}}^{(5)}_{\theta}\right)
      - \frac{2\pi G}{3\rho_{\Lambda}}\left[\rho\left(2\rho + 3p\right) \right.\nb\\
      & &  
      \left. + \left(\rho + 3p
       - \tau_{(\phi, \psi)}\right) \tau_{(\phi, \psi)}\right],
\eqn 
where $H \equiv \dot{a}/{a}, \; \rho_{\Lambda} \equiv {\Lambda_{4}}/({8\pi G_{4}})$, and
 \bqn
 \lb{4.13}
 {\cal{G}}^{(5)}_{\tau} &\equiv& \frac{1}{3}e^{-2\sigma}\left[\left({\phi_{,t}}^{2} + {\psi_{,t}}^{2}\right)
 - \left({\phi_{,y}}^{2} + {\psi_{,y}}^{2}\right)\right]\nb\\
 & & - \frac{1}{24}\left\{5\left[\left(\nabla\phi\right)^{2}  + \left(\nabla\psi\right)^{2}\right]
 - 6 V_{5}\right\},\nb\\
 {\cal{G}}^{(5)}_{\theta} &\equiv&  \frac{1}{24}\left\{ 8 \left({\phi_{,n}}^{2} + {\psi_{,n}}^{2}\right)
 - 6 V_{5}\right.\nb\\
 & & \left. + 5\left[\left(\nabla\phi\right)^{2} + \left(\nabla\psi\right)^{2}\right]\right\},\nb\\
 E^{(5)} &\equiv& \frac{1}{6}e^{-2\sigma}\left[\left(\sigma_{,tt} - \omega_{,tt}\right)
 - \left(\sigma_{,yy} - \omega_{,yy}\right) \right.\nb\\
 & & \left. + k e^{2(\sigma - \omega)}\right],
 \eqn
with $\phi_{,n} \equiv n^{a}\nabla_{a}\phi$.   
If the typical size of the extra dimensions is $R$, then it can be shown that
\bq
\lb{4.16}
\rho_{\Lambda} = \frac{\Lambda_{4}}{8\pi G_{4}} = 
3 \left(\frac{R}{{{l}}_{pl}}\right)^{10}\left(\frac{M_{10}}{M_{pl}}\right)^{16}
{M_{pl}}^{4},
\eq
where  ${M_{pl}}$ and ${{l}}_{pl}$ denote the Planck mass and length,  respectively. 
If $M_{10}$ is in the order of TeV  \cite{Anto00}, we find that, in order to have
$\rho_{\Lambda}$ be in the order of its  current observations value $\rho_{\Lambda} 
\simeq 10^{-47}\; GeV^{4}$,  the typical size of the extra dimensions should be
$R \simeq 10^{-22} \; m$, which is well below the current experimental limit of 
the extra dimensions  \cite{Hoyle}.

\section{A Particular Model }
\renewcommand{\theequation}{3.\arabic{equation}}
\setcounter{equation}{0} 

In this section, we    consider a specific solution of the five-dimensional
bulk and the corresponding Friedmann equations on the orbifold branes.

\subsection{Exact Solutions in the Bulk}

It can be shown that the following solution satisfies the  
field equations in the bulk,  
\bqn
\lb{5.1}
\sigma(t) &=& \frac{1}{9}\;\ln(t) + \frac{1}{2}\ln\left(\frac{7}{6}\right),\nb\\
\omega(t) &=& \frac{10}{9}\; \ln(t),\nb\\
\phi(t) &=& -\frac{5}{18}\sqrt{6}\; \ln(t) + \phi_{0},\nb\\
\psi(t) &=& -\frac{\sqrt{10}}{6}\; \ln(t) + \psi_{0},
\eqn
for $k = -1$, where
\bq
\lb{5.2}
\phi_{0}  = \frac{\sqrt{6}}{5}\left\{\ln\left(\frac{2}{3V^{0}_{(5)}}\right)
- \sqrt{\frac{5}{2}}\; \psi_{0}\right\},
\eq
with $\psi_{0}$ being an arbitrary constant. Then, the corresponding 5-dimensional metric takes
the form,
\bq
\lb{5.3}
ds^{2}_{5} = \left(\frac{7}{6}\right)t^{2/9}\left(dt^{2} - dy^{2}\right)
- t^{20/9}d\Sigma^{2}_{-1}.
\eq
Clearly, the spacetime is singular at $t = 0$ where all the four spatial dimensions collapse into
a point singularity, a big bang like. This can be seen more clearly from the expression,
\bq
\lb{5.4}
\psi_{,a}\psi^{,a} = \frac{3}{5}\phi_{,a}\phi^{,a} = \frac{5}{21} t^{-20/9}.
\eq
The corresponding Penrose diagram is given by Fig. 1. 

\begin{figure}
\centering
\includegraphics[width=8cm]{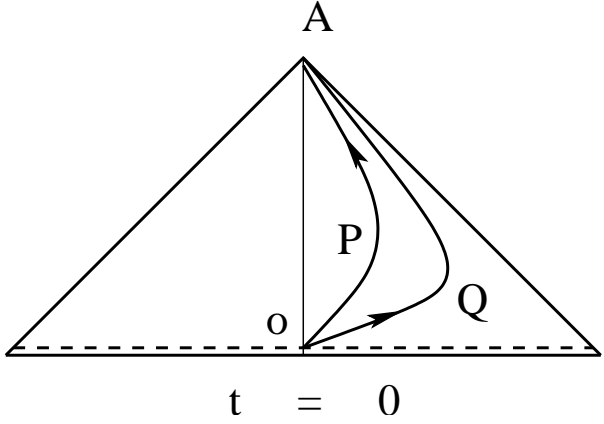}
\caption{The Penrose diagram for the metric given by
Eq.(\ref{5.3}) in the text, where the spacetime is singular at $t = 0$.  The curves $OPA$
and $OQA$ describes the history of the two orbifold  branes located on the surfaces
$y = y_{I}(\tau_{I})$ with $I = 1, 2$. The bulk is the region between these two lines. }
\label{fig1}
\end{figure}

Lifting the solution to the 10-dimensional superstring spacetime, we find that in the string
frame the metric (\ref{2.3}) takes the form,
\bqn
\lb{5.5}
d{\hat{s}}^{2}_{10} &=& \hat{g}_{AB} dx^{A}dx^{B} \nb\\
&=& e^{\sqrt{\frac{2}{3}}\; \phi_{0}}\left\{\left(\frac{7}{6}\right)t^{-1/3}\left(dt^{2} - dy^{2}\right)
- t^{5/3}d\Sigma_{-1}^{2}\right\}\nb\\
& & - e^{\sqrt{\frac{2}{5}}\; \psi_{0}} t^{-1/3}\delta_{ij}dz^{i}dz^{j}.
\eqn
The corresponding dilaton field is given by
\bq
\lb{5.6}
\hat{\Phi}  =  - \frac{5}{3}\ln(t) + \hat{\Phi}_{0},
\eq
where $\hat{\Phi}_{0} \equiv \sqrt{3/2}\; \phi_{0} +  \sqrt{5/2}\; \psi_{0}$, from which we find
\bq
\lb{5.7}
\hat{\Phi}_{,A}\hat{\Phi}^{,A} = \frac{50}{21}e^{-\sqrt{\frac{2}{3}}\; \phi_{0}}
t^{-5/3}.
\eq
Clearly, it is also singular at $t = 0$, but with a weaker strength in comparing to that of the 
five-dimensional spacetime given by Eq.(\ref{2.17}). A critical difference is that in the 
string frame the proper distance along the $y$-direction becomes decreasing as $t$ increases,
in contrast to that in the Einstein frame, as can be seen clearly from Eqs.(\ref{5.3}) and
(\ref{5.5}).

\subsection{Generalized Friedmann Equations on The Branes}

On the other hand, from Eq.(\ref{4.13})   we find that
\bqn
\lb{5.10}
E^{(5)} &=& - \frac{1}{42 a^{2}}, \;\;\; 
{\cal{G}}^{(5)}_{\tau} = \frac{31}{126 a^{2}},\nb\\
{\cal{G}}^{(5)}_{\theta} &=& \frac{20}{81 a^{9/5}} {\dot{y}}^{2} -
\frac{13}{378 a^{2}},
\eqn
where now $a(\tau) = t^{10/9}(\tau)$, and $\dot{y}$ is given by
\bqn
\lb{5.11}
\dot{y} &=& \epsilon_{y}a^{9/10} \left[\left(\frac{9}{10}\right)^{2}H^{2}
-    \frac{6}{7a^{2}}\right]^{1/2},
\eqn
with $\epsilon_{y} = \pm 1$. Inserting Eqs.(\ref{5.10}) and (\ref{5.11})
into Eqs.(\ref{4.14a}) and (\ref{4.14b}), we find that
\bqn
\lb{5.12a}
H^{2}   &=& \frac{8\pi G}{3}\left(\rho + \tau_{(\phi, \psi)} + \rho_{\Lambda}\right)
 + \frac{200}{189 a^{2}} \nb\\
& & + \frac{2\pi G}{3\rho_{\Lambda}}\left(\rho + \tau_{(\phi, \psi)}\right)^{2},\\
\lb{5.12b}
\frac{\ddot{a}}{a}   &=&  \frac{4\pi G}{5}\left(3\rho_{\Lambda} +3 \tau_{(\phi, \psi)}
-2\rho - 5p\right)  
      \nb\\
      & &  - \frac{2\pi G}{3\rho_{\Lambda}}
      \left[\frac{1}{10}\left(\rho + \tau_{(\phi, \psi)}\right)^{2}
      +\rho\left(2\rho + 3p\right)\right. \nb\\
      & &  
      \left. + \left(\rho + 3p
       - \tau_{(\phi, \psi)}\right) \tau_{(\phi, \psi)}\right].
\eqn 
It is remarkable to note that these two equations do not depend on both $\epsilon$ defined in
Eq.(\ref{4.11}) and $\epsilon_{y}$ defined in
Eq.(\ref{5.11}). Combining Eqs.(\ref{5.12a}) and (\ref{5.12b}), we obtain
\bqn
\lb{5.12c}
\left(\dot{\rho} + \dot{\tau}_{(\phi, \psi)}\right) + 3H\left(\rho + p\right)
&=& - \frac{H}{20\Delta}\left[4\left(\rho + \rho_{\Lambda} + \tau_{(\phi, \psi)}\right)
   \right.\nb\\
   & & \left.
   + \frac{\left(\rho + \tau_{(\phi, \psi)}\right)^{2}}{\rho_{\Lambda}}\right],
\eqn
where
\bq
\lb{5.13}
\Delta  \equiv   1 + \frac{1}{2\rho_{\Lambda}}\left(\rho + \tau_{(\phi, \psi)}\right).
\eq
Eq.(\ref{5.12c}) shows clearly the interaction among the matter fields confined on the
branes and the bulk. This can also be seen from Eq.(\ref{5.10}).

\subsection{Current Acceleration of the Universe}
To study current acceleration of the universe, we first set 
\bq
\lb{5.14a}
p = 0,
\eq
and then introduce the quantities,
\bqn
\lb{5.14}
\Omega_{m} &=& \frac{\rho_{m}}{\rho_{cr}}, \;\;\;
\Omega_{\tau} = \frac{\tau_{(\phi,\psi)}}{\rho_{cr}},\;\;\;
\Omega_{\Lambda} = \frac{\rho_{\Lambda}}{\rho_{cr}},\nb\\ 
\Omega_{k} &=& \frac{200}{189 H^{2}_{0} a^{2}} =  \frac{\Omega_{k}^{(0)}}{a^{2}}, 
\eqn
where  $\rho_{cr} \equiv 3H^{2}_{0}/8\pi G$. It should be noted the slight difference
between $\Omega_{k}$ defined here and the one normally used, 
$\Omega_{k} = - {k}/({H^{2}_{0} a^{2}})$. Then,  Eqs.(\ref{5.12a}), (\ref{5.12c}) 
and (\ref{5.11}) can be written as
\bqn
\lb{5.15a}
E^{2} &=& \Omega_{\Lambda} + \Omega_{t} + \Omega_{k} 
           + \frac{{\Omega_{t}}^{2}}{4\Omega_{\Lambda}},\\
\lb{5.15b}
\Omega_{t}^{*} &=& - \frac{E}{\Delta}\left\{\frac{1}{5}\left(\Omega_{\Lambda}
                    + 16 \Omega_{t} - 15\Omega_{\tau}\right)\right.\nb\\
	& & \left. + \frac{\Omega_{t}}{20\Omega_{\Lambda}}
	     \left(31\Omega_{t} - 30\Omega_{\tau}\right)\right\},\\
\lb{5.15c}
y^{*} &=&  \epsilon_{y}\left(\frac{9}{10}\right)
     \left(\frac{\Omega_{k}^{(0)}}{\Omega_{k}}\right)^{9/20}\nb\\
     & & \times \sqrt{\Omega_{\Lambda} + \Omega_{t}   
           + \frac{{\Omega_{t}}^{2}}{4\Omega_{\Lambda}}}, 
\eqn
where $E \equiv H/H_{0}, \; y^{*} \equiv dy/d(H_{0}\tau)$, and
\bq
\lb{5.16}
\Omega_{t} = \Omega_{m} + \Omega_{\tau},
\eq
with  the constraint,
\bq
\lb{5.17}
 1 = \Omega_{k}^{(0)} + \Omega_{\Lambda} + \Omega_{t}^{(0)} 
+ \frac{{\Omega_{t}^{(0)}}^{2}}{4\Omega_{\Lambda}},
\eq
where $\Omega_{N}^{(0)}$'s denote their current values. On the other hand, 
in terms of $\Omega$'s, we find
\bqn
\lb{5.19}
\frac{a^{**}}{a} &=& \frac{3}{10}\left(3\Omega_{\Lambda} - 2\Omega_{t}
+ 5\Omega_{\tau}\right) \nb\\
& &  + \frac{3\Omega_{t}}{40\Omega_{\Lambda}}\left(7\Omega_{t} 
- 10\Omega_{\tau}\right).
\eqn
To study 
Eqs.(\ref{5.15a})-(\ref{5.15c}) and (\ref{5.19}) further, we need to specify 
$\Omega_{\tau}$. In the following, we shall consider two different cases.

\subsubsection{$V_{4}^{(I)} = V_{(4)}^{0} \exp\left\{\frac{n}{2}
                \left(\frac{5}{\sqrt{6}}\; \phi 
		+ \sqrt{\frac{5}{2}}\; \psi\right)\right\}$}

If we choose the potential $V_{4}^{(I)}(\phi, \psi)$ on each of the two branes as
[cf. Eq.(\ref{2.17})],
\bq
\lb{5.17aa}
V_{4}^{(I)} = V_{(4)}^{0} \exp\left\{\frac{n}{2}
\left(\frac{5}{\sqrt{6}}\; \phi + \sqrt{\frac{5}{2}}\; \psi\right)\right\},
\eq
where $V_{(4)}^{0}$ and $n$ are arbitrary constants, we find that
\bq
\lb{5.17a}
\Omega_{\tau}  =  \epsilon_{I}\frac{V_{(4)}^{0}}{\rho_{cr}}
\left(\frac{2}{3 V_{(5)}^{0}}\right)^{n/2} \frac{1}{a^{n}} \equiv
\frac{\Omega_{\tau}^{(0)}}{a^{n}}.
\eq
Then,  our fitting parameters in this case can be chosen as  
\bq
\lb{5.18}
\left\{\Omega_{\Lambda}, \Omega_{m}^{(0)}, \Omega_{k}^{(0)}\right\},
\eq
for any given $n$. 

Fitting the above model to the 182 gold supernova Ia data \cite{Riess06} and the BAO parameter
from SDSS \cite{SDSS}, by using our
numerical code \cite{Gong}, based on the publicly available MINUIT program of CERN, we find that, 
for $n =1$, the best fitting is $\Omega_{m} = 0.24 \pm {}^{0.03}_{0.03}$, 
$\Omega_{\Lambda} = 0.76 \pm {}^{0.37}_{0.27}$, and $\Omega_{k} = 0.00 \pm {}^{0.05}_{0.00}$
with $\chi^{2} = 172.4$. Figs. 2-4 show the marginalized contours of the $\Omega$'s, from which
we can see that the effect of the interaction between the bulk and the brane
is negligible, and the later evolution of the universe follows more or less the same pattern 
as that of the $\Lambda$CDM model in the Einstein theory of gravity.

\begin{figure}
\centering
\includegraphics[width=8cm]{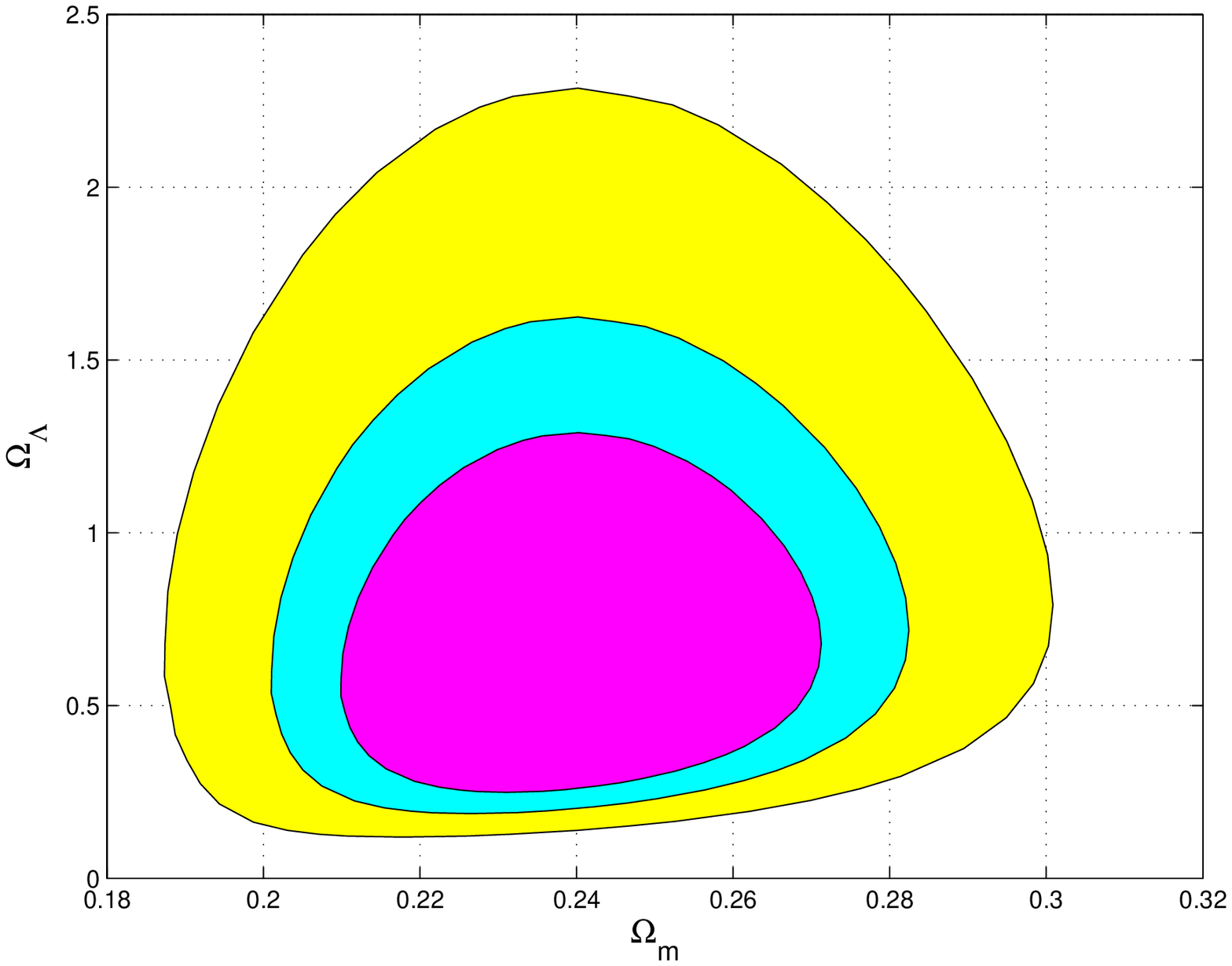}
\caption{The marginalized contour of $\Omega_{m}-\Omega_{\Lambda}$ for the potential given by
Eq.(\ref{5.17aa}) with $n = 1$. }
\label{fig2}
\end{figure}

\begin{figure}
\centering
\includegraphics[width=8cm]{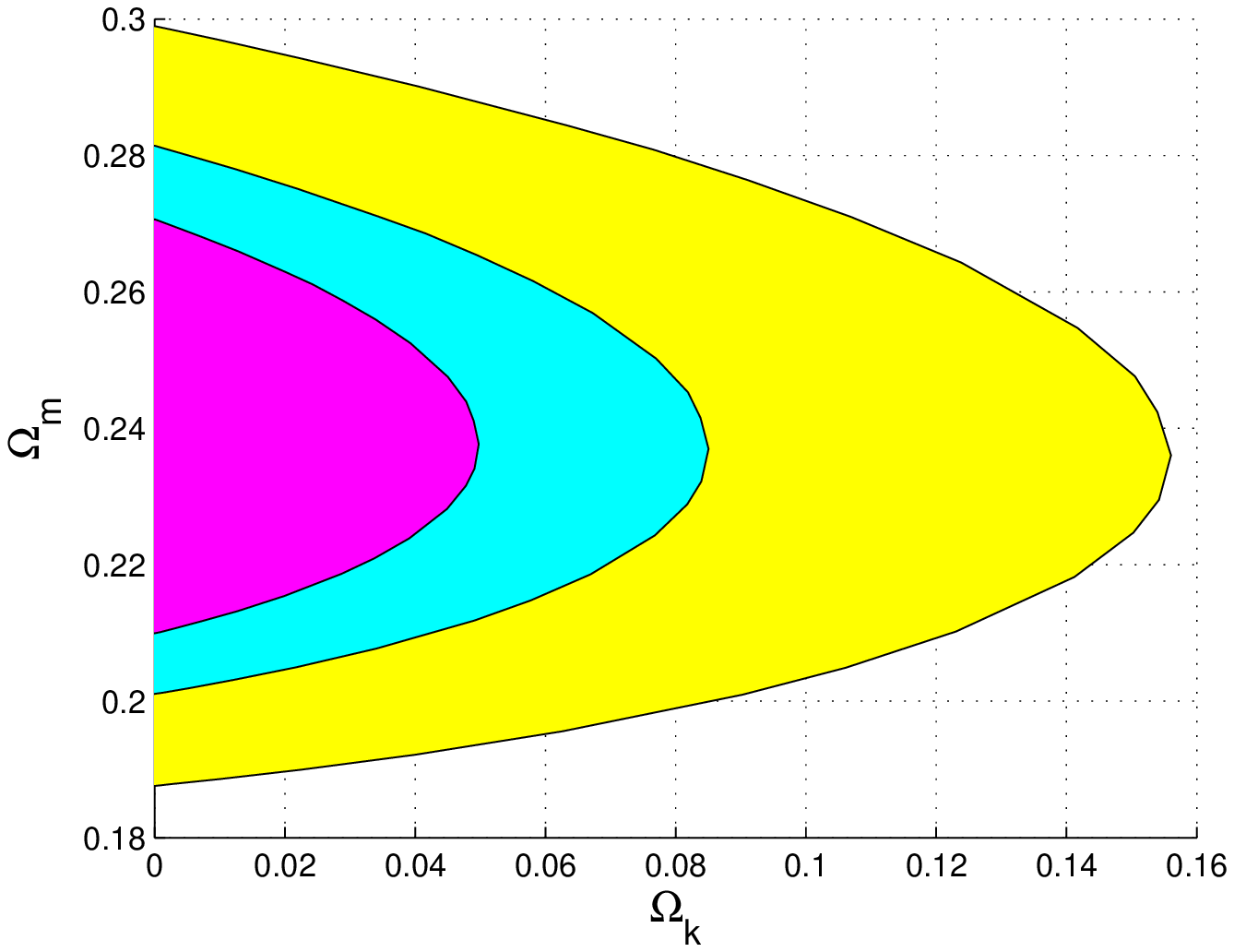}
\caption{The marginalized contour of $\Omega_{m}-\Omega_{k}$ for the potential given by
Eq.(\ref{5.17aa}) with $n = 1$. }
\label{fig3}
\end{figure}

\begin{figure}
\centering
\includegraphics[width=8cm]{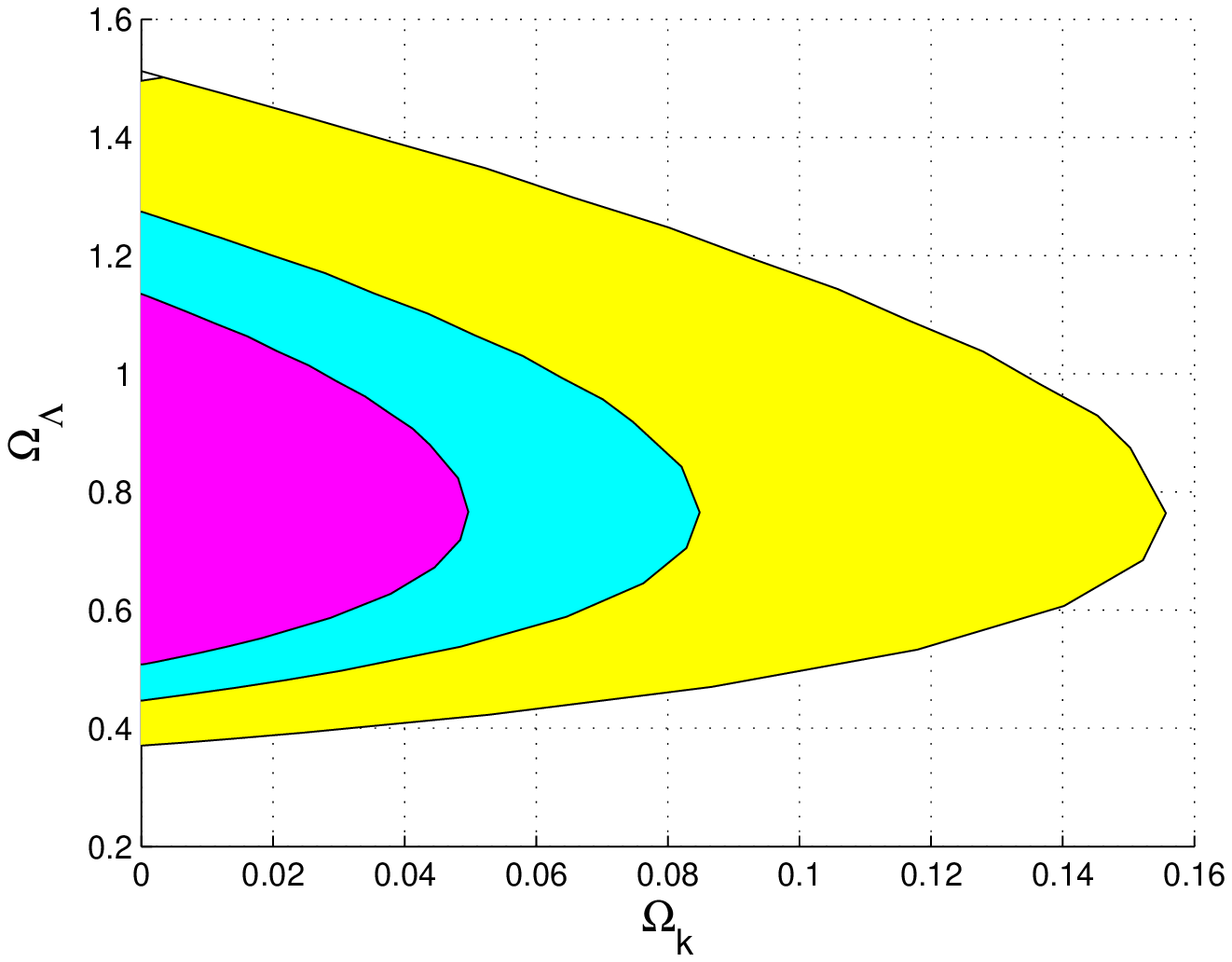}
\caption{The marginalized contour of $\Omega_{k}-\Omega_{\Lambda}$ for the potential given by
Eq.(\ref{5.17aa}) with $n = 1$. }
\label{fig4}
\end{figure}

For $n =3.5$, we find that the best fitting is $\Omega_{m} = 0.27 \pm {}^{0.03}_{0.03}$, 
$\Omega_{\Lambda} = 0.58 \pm {}^{0.11}_{0.12}$, and $\Omega_{k} = 0.00 \pm {}^{0.06}_{0.00}$
with $\chi^{2} = 164.2$. Figs. 5-7 show the marginalized contours of the $\Omega$'s. 

\begin{figure}
\centering
\includegraphics[width=8cm]{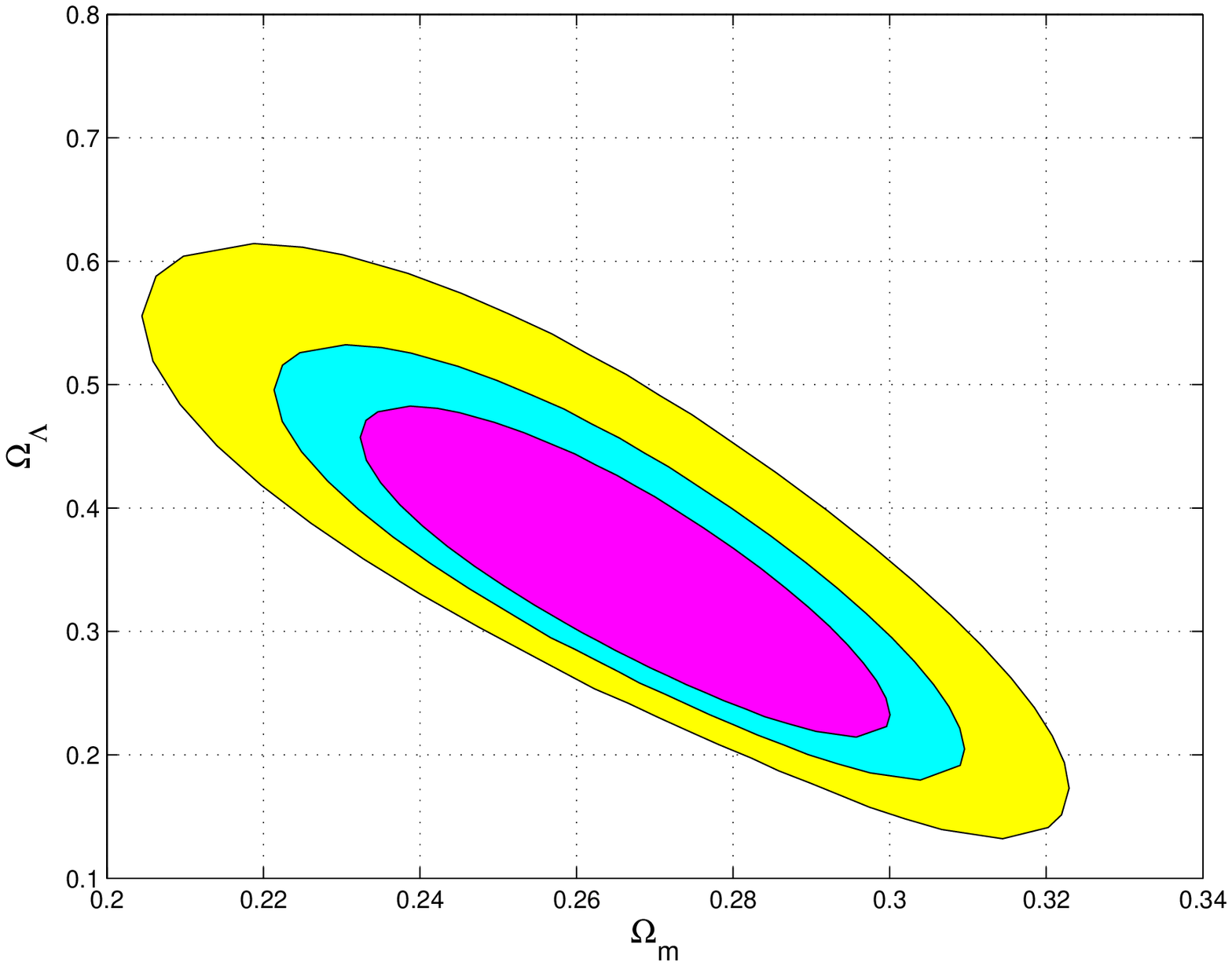}
\caption{The marginalized contour of $\Omega_{m}-\Omega_{\Lambda}$ for the potential given by
Eq.(\ref{5.17aa}) with $n = 3.5$. }
\label{fig5}
\end{figure}

\begin{figure}
\centering
\includegraphics[width=8cm]{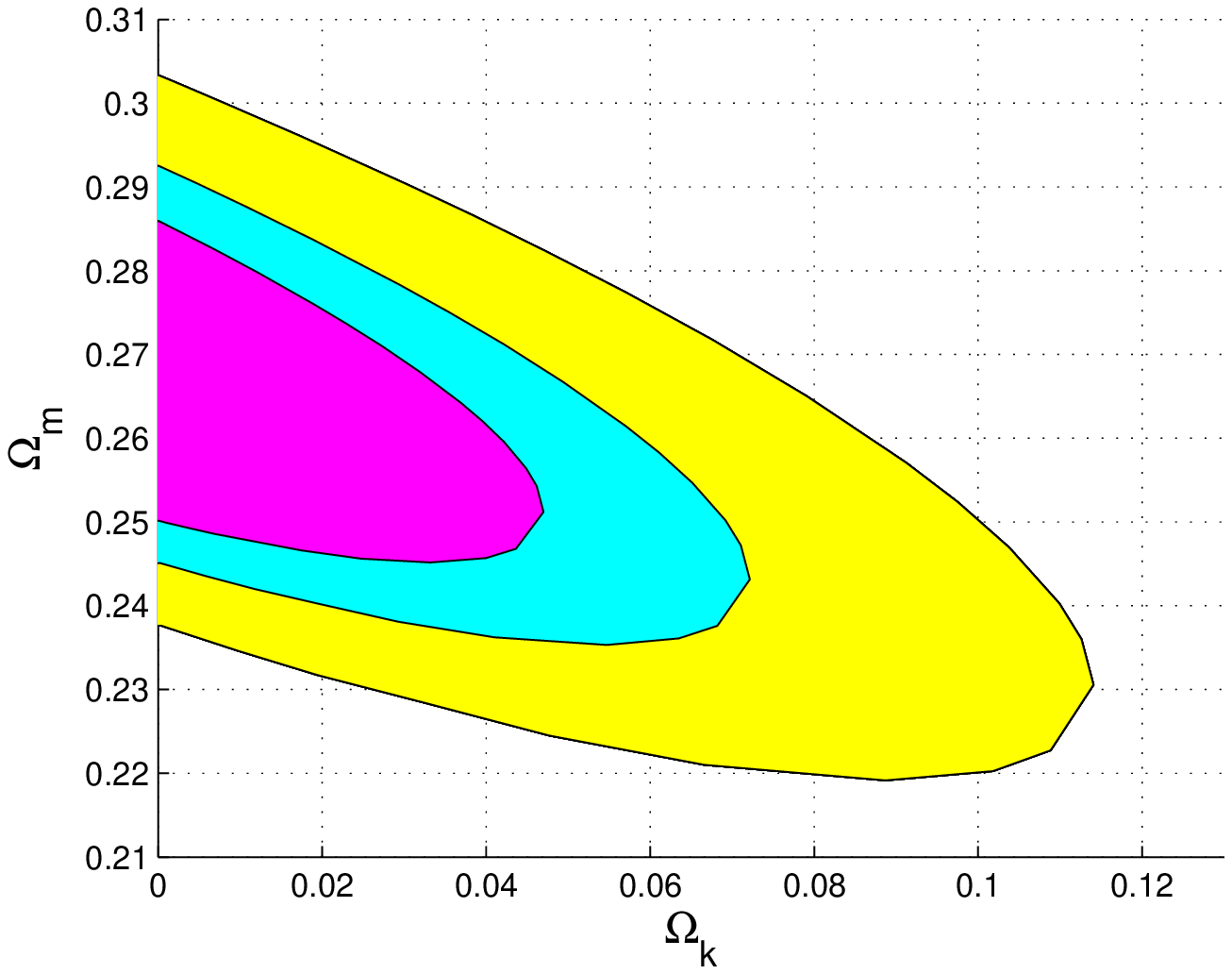}
\caption{The marginalized contour of $\Omega_{m}-\Omega_{k}$ for  the potential given by
Eq.(\ref{5.17aa}) with $n = 3.5$. }
\label{fig6}
\end{figure}

\begin{figure}
\centering
\includegraphics[width=8cm]{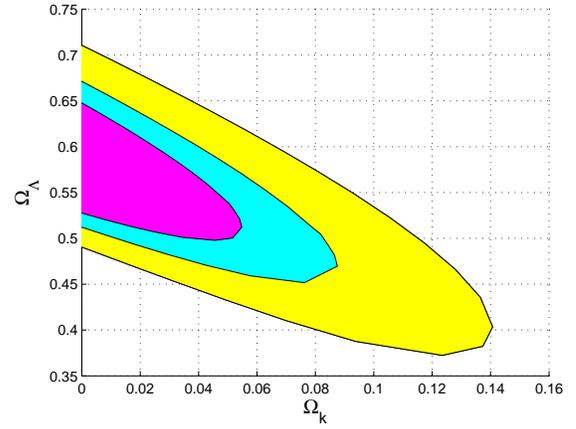}
\caption{The marginalized contour of $\Omega_{k}-\Omega_{\Lambda}$ for the potential given by
Eq.(\ref{5.17aa}) with $n = 3.5$. }
\label{fig7}
\end{figure}

The above shows clearly that the case with $n = 3.5$ is observationally more favorable than that
of $n = 1$.  We have also fitted the data with various  values of $n$, and found  
that the best fitting value of $n$ is about $n = 3.5$.

With the above best fitting values of the $\Omega$'s and $n$ as initial conditions, the future 
evolution of the universe is shown in Figs. 8 and 9, from which we can see that all of them, 
except for $\Omega_{\Lambda}$, decreases rapidly, and   $\Omega_{\Lambda}$ soon dominates
the evolution of the universe, whereby a de Sitter universe is resulted.

\begin{figure}
\centering
\includegraphics[width=8cm]{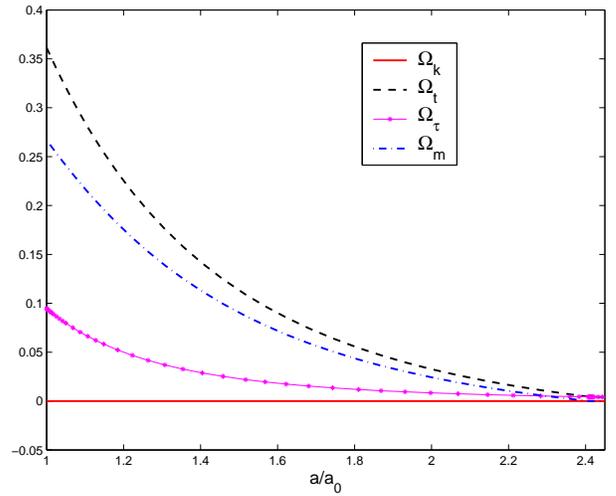}
\caption{The evolution of the matter components, $\Omega_{i}$'s,  for the potential given by
Eq.(\ref{5.17aa}) with $n = 3.5$. }
\label{fig8}
\end{figure}

\begin{figure}
\centering
\includegraphics[width=8cm]{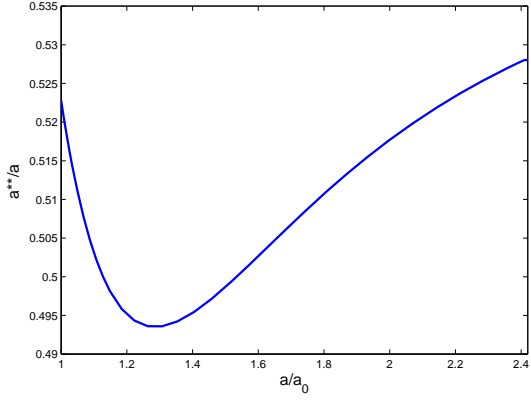}
\caption{The evolution of the acceleration $a^{**}/a \equiv (d^{2}a/d\left(H_{0}\tau\right)^{2})/a$
   for  the potential given by
Eq.(\ref{5.17aa}) with $n = 3.5$. }
\label{fig9}
\end{figure}

\begin{figure}
\centering
\includegraphics[width=8cm]{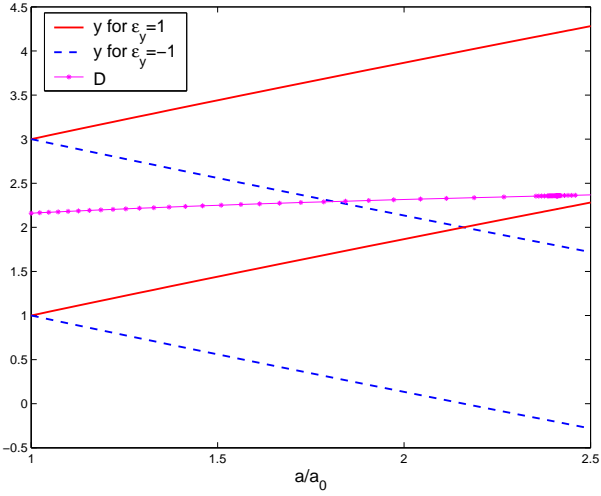}
\caption{The locations of the two branes $y_{I}(a)$, and the proper distance
${\cal{D}}$ between the two branes  for the potential given by
Eq.(\ref{5.17aa}) with $n = 3.5$. The initial conditions are chosen
so that $y_{1}(a_{0}) = 3$ and  $y_{2}(a_{0}) = 1$. The choice of $\epsilon_{y} = + 1$ 
($\epsilon_{y} = -1$) corresponds to the case where the branes move towards the increasing  
(decreasing) direction of $y$.}
\label{fig10}
\end{figure}

From the metrics of Eqs.(\ref{5.3}) and (\ref{5.5}), on the other hand,  one may naively conclude 
that the radion in the present case is not stable, as the proper
distance given by Eq.(\ref{4.9a}) seems either to increases  to infinity (in the Einstein frame, 
given by Eq.(\ref{5.3})) or to decreases to zero 
(in the string frame, given by Eq.(\ref{5.5})), as $t \rightarrow \infty$. A closer investigation 
shows that the problem is not as simple as it looks like. In particular, since $y_{I} = y_{I}(\tau_{I})$,
Eq.(\ref{4.9a}) makes sense only when the relation $\tau_{1} = \tau_{1}(\tau_{2})$ is known. 
In the present case, we transform such a dependence to the expansion factor $a$, and plot it out
in Fig.  \ref{fig10}, together with $y_{I}(a)$, from which we can see clearly that  the distance between the 
two branes remains almost   constant. This indicates that the radion might be stable. 
Certainly, before a definitive conclusion is reached, more detailed investigations are needed.

We also fit the above model with $n = 3.5$ by using our Monte-Carlo Markov Chain (MCMC) code 
\cite{GWW07b}, based on the publicly available package COSMOMC \cite{LB02}, and find that 
the best fitting is $\Omega_{m} = 0.27 \pm {}^{0.04}_{0.03}$, 
$\Omega_{\Lambda} = 0.61 \pm {}^{0.09}_{0.10}$, and $\tilde{\Omega}_{k} = -0.0026 
\pm {}^{0.2339}_{0.2396}$ with $\chi^{2} = 164.10$, where 
\bq
\lb{5.18aa}
{\Omega}_{k} 
\equiv {\tilde{\Omega}_{k}}^{2}.
\eq

\begin{figure}
\centering
\includegraphics[width=8cm]{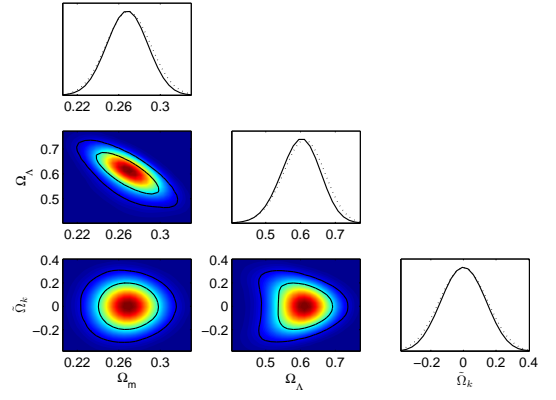}
\caption{The marginalized probabilities and contours for the potential given by
Eq.(\ref{5.17aa}) with $n = 3.5$. }
\label{MCMC}
\end{figure}

The corresponding marginalized probabilities and contours are given in Fig. \ref{MCMC}.
Clearly, these best fitting values are consistent with those obtained above by using 
our MINUIT code \cite{Gong}. 

\subsubsection{$V_{4}^{(I)} = \lambda_{4}^{(I)}  \left(\psi^{2} - {v_{I}}^{2}\right)^{2}$}

To stabilize the radion, Goldberger and Wise proposed to choose the potential 
$V_{4}^{(I)}$ as \cite{GW99},
\bq
\lb{5.20}
V_{4}^{(I)}(\phi, \psi) = \lambda_{4}^{(I)}\left(\psi^{2} - {v_{I}}^{2}\right)^{2},
\eq
where $\lambda_{4}^{(I)}$  and ${v_{I}}^{2}\; (I = 1, 2)$ are constants. Then,  we find that
\bq
\lb{5.21}
\Omega_{\tau}^{(I)} =  \Omega_{\tau}^{(0, I)}\left(\left(\frac{3}{\sqrt{40}}\ln(a)\right)^{2} 
- {v_{I}}^{2}\right)^{2},
\eq
where $\Omega_{\tau}^{(0,I)} \equiv \epsilon_{I} \lambda_{4}^{(I)}/\rho_{cr}$. Note that 
in writing the above expressions,  without loss of any generality, we had set $\psi_{0} = 0$. 
Then, the fitting parameters now can be taken as,
\bq
\lb{5.18a}
\left\{\Omega_{\Lambda}, \Omega_{m}^{(0)}, \Omega_{k}^{(0)}, v_{I}\right\}.
\eq
Fitting the above model to the 182 gold supernova Ia data 
\cite{Riess06} and the BAO parameter from SDSS \cite{SDSS}, 
we first study the dependence of $\chi^{2}$ on $v_{I}$. Table \ref{table1}
shows such a dependence and the best fitting values of $\Omega_{i}$'s for each given 
$v_{I}$. 
\begin{table}
\begin{tabular}{|c|c|c|c|c|c|}
\hline
$v_{I}$ & $\chi^{2}$ & $\Omega_{m}$  & $\tilde{\Omega}_{k}$ & $\Omega_{\Lambda}$ 
& $\Omega_{\Lambda} + \Omega_{\tau}$ \\ \hline
$0.1$ & $171.28$ & $ 0.25 \pm^{0.03}_{0.04}$   & $-0.0009 \pm^{0.21}_{0.22}$
& $ 0.72 \pm^{0.05}_{0.05}$ & $0.73$\\ \hline
$0.3$ & $168.10$ & $ 0.29 \pm^{0.05}_{0.05}$   & $-0.0006 \pm^{0.41}_{0.41}$
& $ 1.06 \pm^{0.15}_{0.17}$ & $0.47$\\ \hline
$0.5$ & $157.50 $ & $ 0.29 \pm^{0.04}_{0.04}$   & $-0.008 \pm^{0.46}_{0.44}$
& $ 1.28 \pm^{0.31}_{0.28}$ & $0.70$\\ \hline
$1.0$ & $156.69 $ & $ 0.29 \pm^{0.03}_{0.04}$   & $-0.002 \pm^{0.52}_{0.52}$
& $ 1.64 \pm^{0.71}_{0.48}$ & $0.64$\\ \hline
$3.0$ & $156.38 $ & $ 0.28 \pm^{0.03}_{0.04}$   & $-0.008 \pm^{0.53}_{0.56}$
& $ 1.93 \pm^{1.01}_{0.73}$ & $0.57$\\ \hline
$10.0$ & $166.35 $ & $ 0.28 \pm^{0.05}_{0.03}$   & $-0.002 \pm^{0.62}_{0.62}$
& $ 1.97 \pm^{2.17}_{0.74}$ & $0.56$\\ \hline
\end{tabular}
\caption{The best fitting values of $\Omega_{i}$ for a given $v_{I}$ of 
the potential given by Eq.(\ref{5.20}).}
\lb{table1}
\end{table}

From the table we can see that $\chi^{2}$ decreases
until  $v_{I} \simeq 3.0$ and then starts to increase, as $v_{I}$ is continuously
increasing. However, $\Omega_{\Lambda}$ and its uncertainty also increase 
as $v_{I}$ is increasing, while $\Omega_{m}$ and $\Omega_{k}$ 
remain almost the same. Since $\Omega_{\tau}$
acts as a varying cosmological constant, Table \ref{table1} shows that
the total effective cosmological constant $\Omega^{eff.}_{\Lambda} \equiv \Omega_{\Lambda} 
+ \Omega_{\tau}$ is between   $0.47$ and $0.73$. 

\begin{figure}
\centering
\includegraphics[width=8cm]{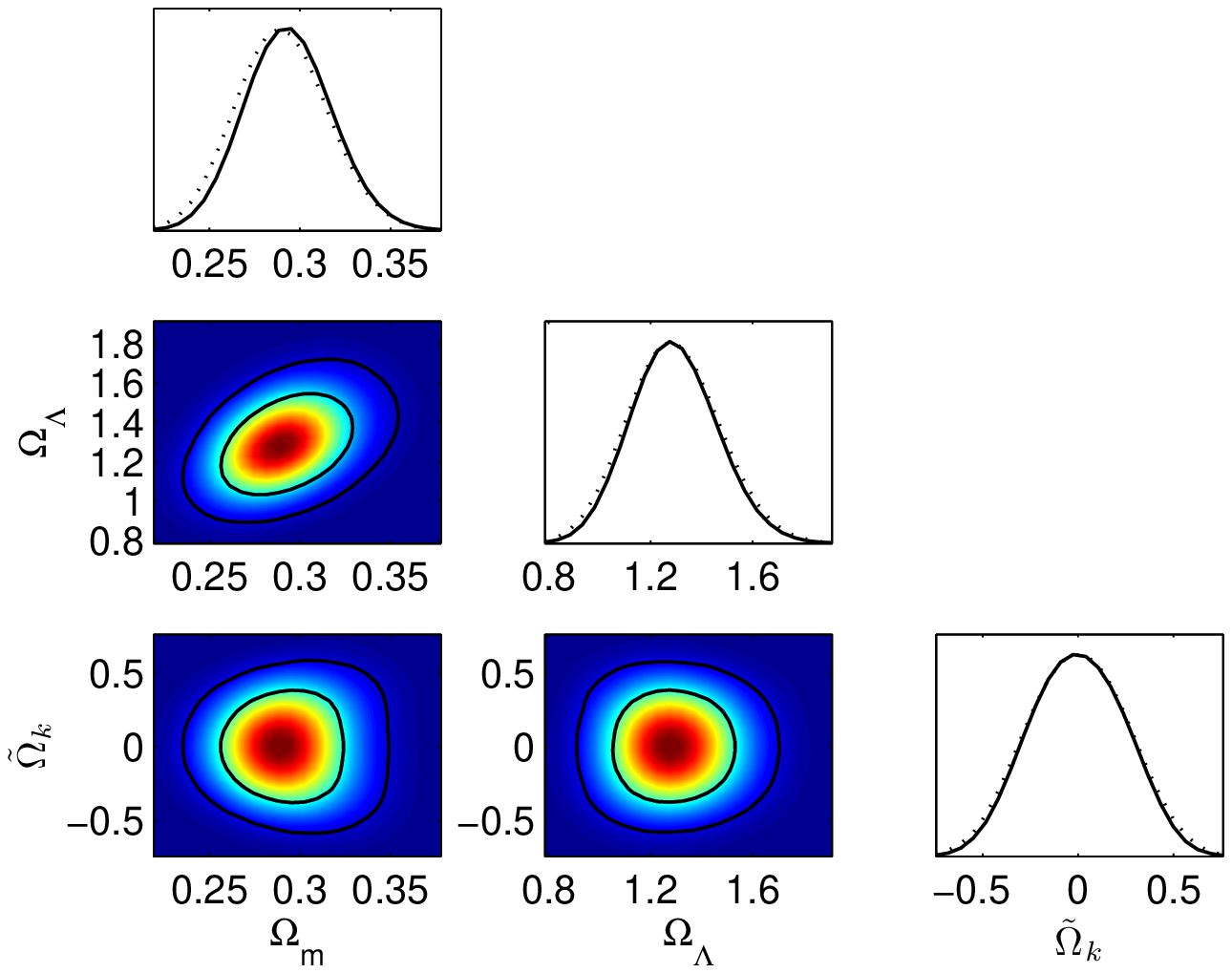}
\caption{The marginalized probabilities and contours for the potential given by
Eq.(\ref{5.20}) with  $v_{I} = 0.5$. }
\label{v05}
\end{figure}

\begin{figure}
\centering
\includegraphics[width=8cm]{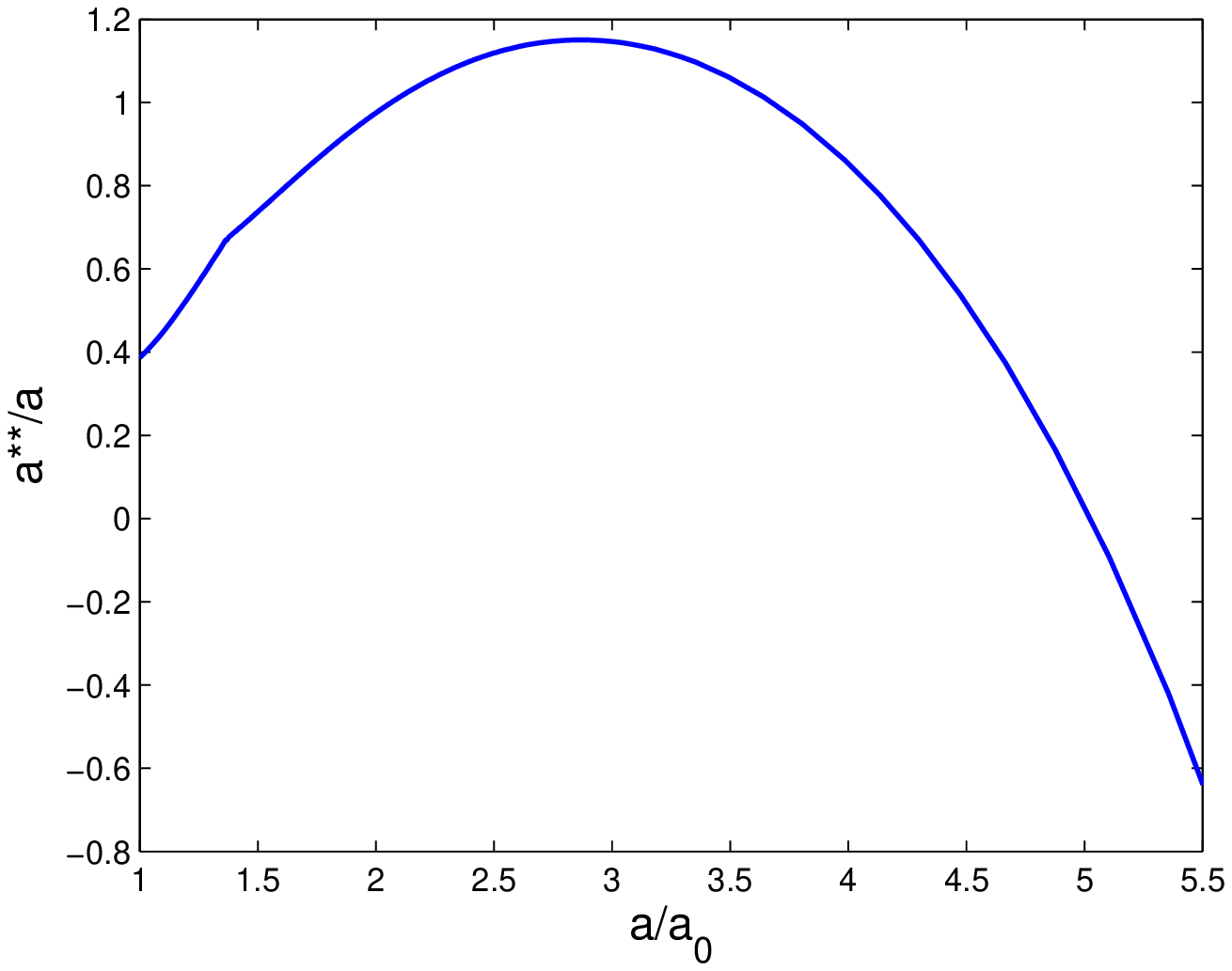}
\caption{The acceleration $a^{**}/a$  for the potential given by
Eq.(\ref{5.20}) with  $v_{I} = 0.5$. }
\label{v05_add}
\end{figure}

\begin{figure}
\centering
\includegraphics[width=8cm]{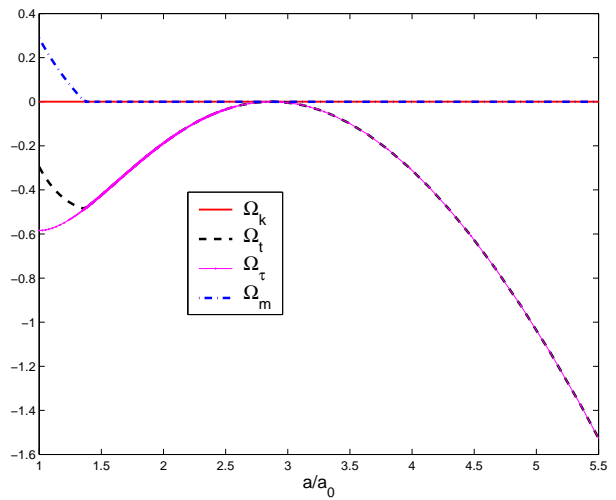}
\caption{The future evolution of $\Omega_{i}$  for the potential given by
Eq.(\ref{5.20}) with  $v_{I} = 0.5$. }
\label{v05_comps}
\end{figure}

Fig. \ref{v05} shows the marginalized probabilities and contours for the 
potential given by Eq.(\ref{5.20}) with  $v_{I} = 0.5$,
and Fig. \ref{v05_add} shows the future evolution of the corresponding acceleration 
of the universe. From there we can see that the acceleration increases to a maximal
value and then starts to decrease. As the time is continuously increasing, it will
pass the zero point and then becomes negative. Thus, in the present model, the domination
of the cosmological constant is  only temporary.  Due to the presence of the potential
term, represented by $\Omega_{\tau}$, the 
universe will be in its decelerating expansion phase again in the future, 
whereby all problems connected with a far future de Sitter universe are resolved
\cite{KS00}. The effects of $\Omega_{\tau}$ can be seen clearly from Fig. \ref{v05_comps},
from which we can see that both  $\Omega_{m}$ and $\Omega_{k}$ decrease rapidly,
and soon $\Omega_{\tau}$ dominates the evolution of the universe.

\begin{figure}
\centering
\includegraphics[width=8cm]{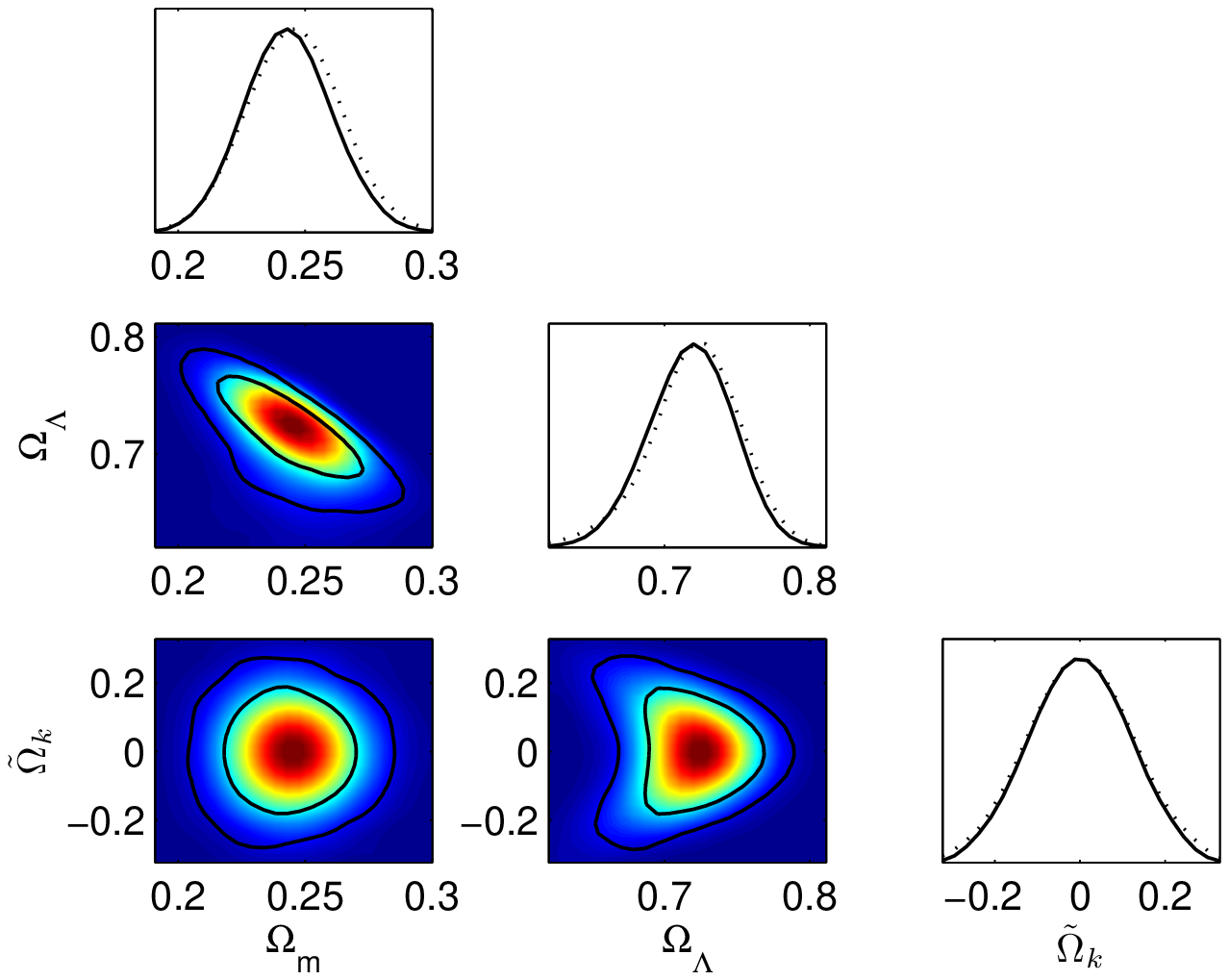}
\caption{The marginalized probabilities and contours for the potential given by
Eq.(\ref{5.20}) with  $v_{I} = 0.1$. }
\label{v01}
\end{figure}

\begin{figure}
\centering
\includegraphics[width=8cm]{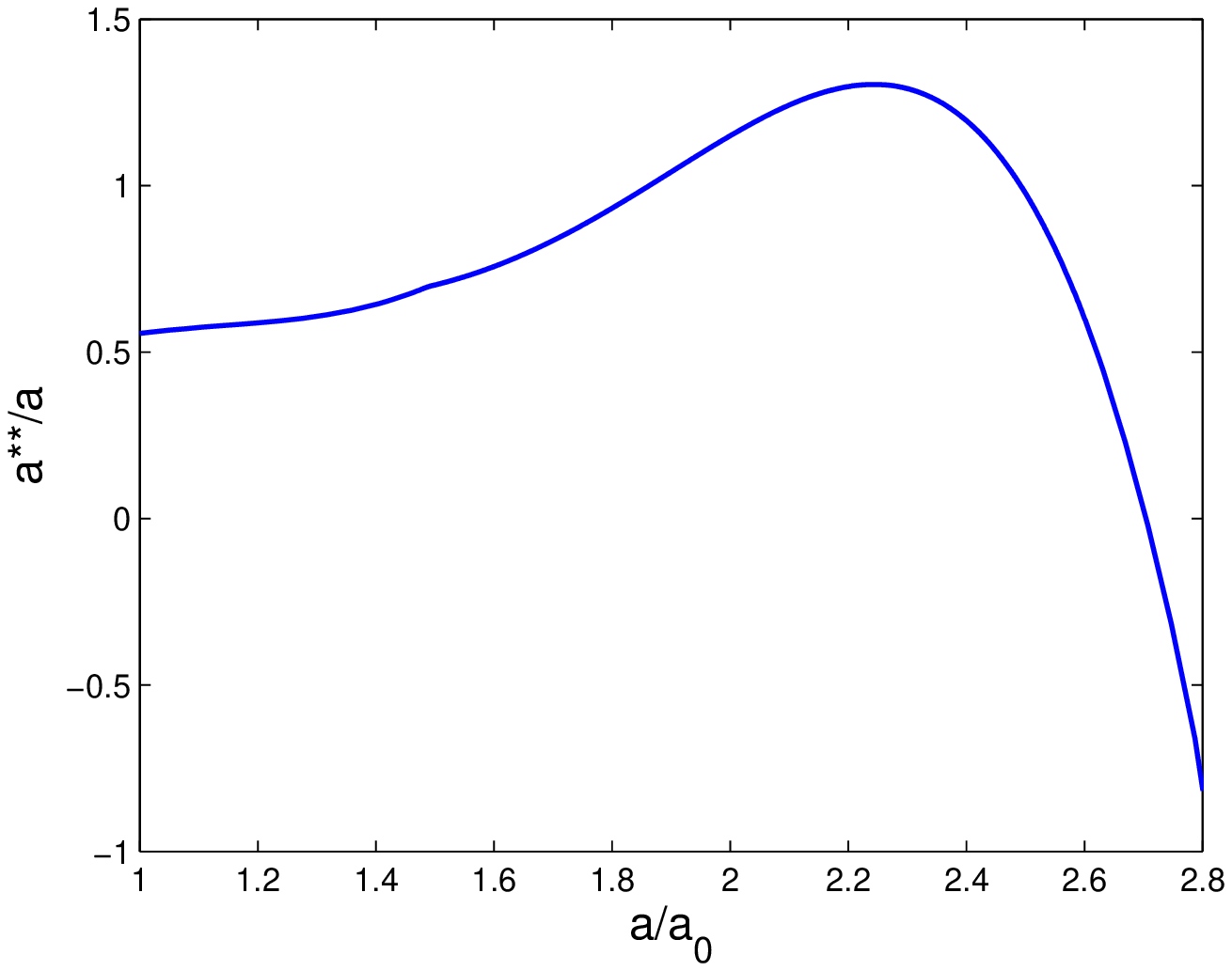}
\caption{The acceleration $a^{**}/a$  for the potential given by
Eq.(\ref{5.20}) with  $v_{I} = 0.1$. }
\label{v01_add}
\end{figure}

\begin{figure}
\centering
\includegraphics[width=8cm]{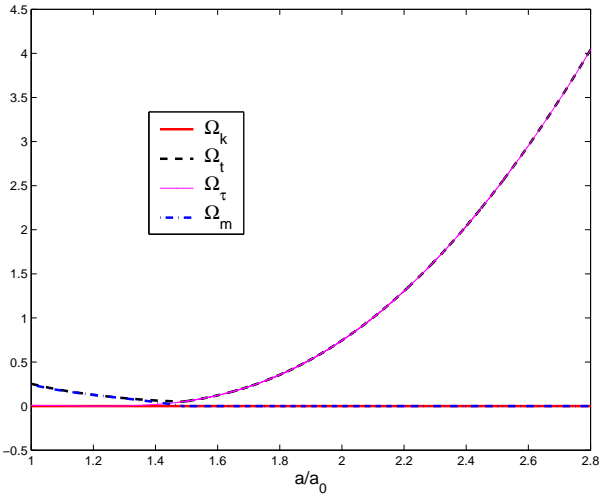}
\caption{The future evolution of $\Omega_{i}$  for the potential given by
Eq.(\ref{5.20}) with  $v_{I} = 0.1$. }
\label{v01_comps}
\end{figure}

These are the common features for any given value of $v_{I}$. Figs. \ref{v01}, \ref{v01_add}, and 
\ref{v01_comps} show, respectively,  the marginalized probabilities and contours,
the future evolution of $a^{**}/a$ and of $\Omega_{i}$ for  $v_{I} = 0.1$.

\begin{figure}
\centering
\includegraphics[width=8cm]{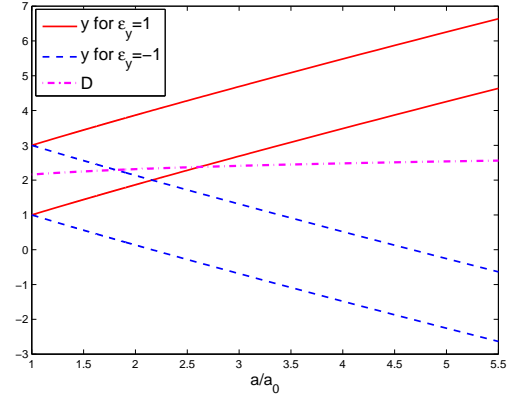}
\caption{The  locations of the two branes, $y_{I}(a)$, and the proper distance, 
${\cal{D}}$, between the two branes    for the potential given by
Eq.(\ref{5.20}) with  $v_{I} = 0.5$. The initial conditions are chosen
so that $y_{1}(a_{0}) = 3$ and  $y_{2}(a_{0}) = 1$. The choice of $\epsilon_{y} = + 1$ ($\epsilon_{y} = -1$)
corresponds to the case where the branes move towards the increasing  (decreasing) direction of $y$.}
\label{v05_yi}
\end{figure}

In addition, we also find that the proper distance between the two orbifold branes defined
by Eq.(\ref{4.9a}) is not sensitive to the choice of $v_{I}$, and remains almost constant
during the future evolution of the universe, as shown in Fig. \ref{v05_yi}. This also
indicates that the radion might be stable in the present case, too. 
 
\section{Conclusions and Remarks}
\renewcommand{\theequation}{4.\arabic{equation}}
\setcounter{equation}{0} 

Recently, we studied the cosmological constant problem in the framework of both  M-theory \cite{GWW07}
and string theory \cite{WS07} on $S^{1}/Z_{2}$, and showed that, among other things, the effective
cosmological constant on the branes can be easily lowered to its current observational value using
the ADD large extra dimension mechanism \cite{ADD98}. Thus, brany cosmology of string/M-Theory seems 
to have a built-in mechanism for solving both the cosmological constant and the hierarchy problems.

In this paper, we have studied a particular cosmological model in the framework of string theory
on $S^{1}/Z_{2}$, developed in \cite{WS07}. We have first solved the field equations in the bulk
and then studied its local and global properties. In particular, we have found that the 10-dimensional
bulk has a big-bang-like singularity at $t = 0$.  

After obtained explicitly the generalized Friedmann equations on each of the two branes 
for any radion potentials $V_{4}^{(I)} \; (I = 1, 2)$ of the branes, we have studied
two different cases where $V_{4}^{(I)} = V_{(4)}^{0} \exp\left\{\frac{n}{2}\left(\frac{5}{\sqrt{6}}\; 
\phi + \sqrt{\frac{5}{2}}\; \psi\right)\right\}$ and $V_{4}^{(I)} = \lambda_{4}^{(I)}
\left(\psi^{2} - {v_{I}}^{2}\right)^{2}$, respectively. For each of these potentials, we have fit 
the corresponding models to the 182 gold supernova Ia data \cite{Riess06} and the BAO parameter
from SDSS \cite{SDSS}, and obtained
the best fitting values of the parameters involved. To doubly check our
numerical codes, we have used both of our  MINUIT  \cite{Gong} and Monte-Carlo Markov Chain (MCMC)  
\cite{GWW07b} codes, and gotten the same results within the allowed errors. With these best 
fitting values as the initial condition, we have integrated numerically the field equations 
on the branes, and found the future evolution of the universe. In particular, for  $V_{4}^{(I)} 
= \lambda_{4}^{(I)}
\left(\psi^{2} - {v_{I}}^{2}\right)^{2}$,  we have found that the current 
acceleration of the universe driven by the effective cosmological constant is only 
temporary.  Due to   the effects of the potentials, the universe will be in its decelerating 
expansion phase again in the future. We have also studied the proper distance between the 
two orbifold branes, and found that it remains almost constant during the whole 
future evolution of the universe in all these models.  

In the framework of orbifold branes, an important question is the radion stability. The considerations
of  the proper distance between the two orbifold branes  indicate that in these cases
the radion might be stable, although further studies are highly demanded. Recently, two of
the authors (NOS $\&$ AW)    studied the problem in a static background with a four-dimensional 
Poincar\'e symmetry,
\bq
\lb{6.2}
ds^{2}_{5} = e^{2\sigma(y)}\left(\eta_{\mu\nu}dx^{\mu}dx^{\nu} - dy^{2}\right),
\eq
where
\bqn
\lb{6.3}
\sigma(y) &=& \frac{1}{9}\ln\left| y + y_{0}\right|,\nb\\
\phi(y) &=& - \frac{5}{\sqrt{54}}\ln\left| y + y_{0}\right| + \phi_{0},\nb\\
\psi(y) &=& - \sqrt{\frac{5}{18}}\ln\left| y + y_{0}\right| + \psi_{0}, 
\eqn
and $y_{0},\; \phi_{0}$ and $\psi_{0}$ are the integration constants with,  
\bq
\lb{6.4}
\psi_{0} = \sqrt{\frac{2}{5}}\left[\ln\left(\frac{2}{9V^{0}_{(5)}}\right) 
- \frac{5}{\sqrt{6}}\phi_{0}\right].
\eq
Following \cite{radion}, we are currently investigating the radion stability, and
wish to report our findings some where else soon.

\section*{Acknowledgments} 

The authors AW and QW would like to thank Dr. Yungui Gong for valuable discussions
and suggestions. This work was initiated when the authors NOS and AW were visiting 
LERMA/CNRS-FRE, Paris. They would like to thank the Laboratory for hospitality. 
This work was partially supported  by  NSFC under grant No. 10775119 and 
No. 10703005 (AW $\&$ QW).

\end{document}